\documentclass[12pt,letterpaper]{article}
\pdfoutput=1

\usepackage{graphicx,array}
\usepackage[dvipsnames]{xcolor}
\definecolor{darkblue}{rgb}{0,0.1,0.5}
\definecolor{darkgreen}{rgb}{0,0.5,0.2}
\definecolor{darkred}{RGB}{153,26,0}
\usepackage{ulem}
\definecolor{seablue}{rgb}{0,0.2,0.6}
\usepackage{latexsym}
\usepackage{amssymb, amsmath}
\usepackage{bm}        % for math
\usepackage[numbers,sort&compress]{natbib}
\usepackage{bm,relsize}
\usepackage{slashed}

\definecolor{viola}{RGB}{134,41,198}
\usepackage{mathrsfs}
\usepackage{hyperref} %Automatically links \label and \ref commands; Always load last
\hypersetup{
    colorlinks=true,       % false: boxed links; true: colored links
    linkcolor=darkblue,          % color of internal links
    citecolor=BrickRed,        % color of links to bibliography
    filecolor=magenta,      % color of file links
    urlcolor=MidnightBlue           % color of external links
}
\usepackage[all]{hypcap} %Link navagates to top of figure instead of caption (below fig)
\usepackage{subcaption}

\usepackage[font=small,labelfont=bf,labelsep=period]{caption}

\newcommand{\GeV}{\mathrm{GeV}}
\newcommand{\TeV}{\mathrm{TeV}}
\newcommand{\keV}{\mathrm{keV}}
\newcommand{\Mpl}{M_{\rm Pl}}

\newcommand{\be}{\begin{equation}}
\newcommand{\ee}{\end{equation}}

\newcommand{\op}{\mathcal{O}}

 \date{\today}

\begin{document}

%%%%%%%%%%%%%%%%%%%%%%%%%%%%%%%%%%%%%%%%%%%%%%%%%%%%%%%%%%%%%%%%%%%%%%%%%%
\begin{flushright}

\end{flushright}
\vspace{.6cm}
\begin{center}
{\huge \bf 
Dark QCD Matters
}\\
\bigskip\vspace{1cm}
{
\large Raghuveer Garani, Michele Redi, Andrea Tesi
}
\\[7mm]
 {\it \small
INFN Sezione di Firenze, Via G. Sansone 1, I-50019 Sesto Fiorentino, Italy\\
Department of Physics and Astronomy, University of Florence, Italy\\
 }

\end{center}
%%%%%%%%%%%%%%%%%%%%%%%%%%%%%%%%%%%%%%%%%%%%%%%%%%%%%%%%%%%%%%%%%%%%%%%%%%

\bigskip \bigskip \bigskip \bigskip

%%%%%%%%%%%%%%%%%%%%%%%%%%%%%%%%%%%%%%%%%%%%%%%%%%%%%%%%%%%%%%%%%%%%%%%%%%
\centerline{\bf Abstract} 
\begin{quote}
We investigate the nightmare scenario of dark sectors that are made of non-abelian gauge theories with fermions, gravitationally coupled to the Standard Model (SM).
While testing these scenarios is experimentally challenging, they are strongly motivated by the accidental stability of dark baryons and pions, that explain the cosmological stability of dark matter (DM).
We study the production of these sectors which are minimally populated through  gravitational freeze-in, leading to a dark sector temperature much lower
than the SM, or through inflaton decay, or renormalizable interactions producing warmer DM.
Despite having only gravitational couplings with the SM these scenarios turn out to be rather predictive depending roughly on three parameters: the dark sector temperature, the confinement scale 
and the dark pion mass. In particular, when the initial temperature is comparable to the SM one these scenarios are very constrained by structure formation, $\Delta N_{\rm eff}$ 
and limits on DM self-interactions. Dark sectors with same temperature or warmer than SM are typically excluded.
\end{quote}
%%%%%%%%%%%%%%%%%%%%%%%%%%%%%%%%%%%%%%%%%%%%%%%%%%%%%%%%%%%%%%%%%%%%%%%%%%

%%%%%%%%%%%%%%%%%%%%%%%%%%%%%%%%%%%%%%%%%%%%%%%%%%%%%%%%%%%%%%%%%%%%%%%%%%
\vfill
\noindent\line(1,0){188}
{\scriptsize{ \\ E-mail:\texttt{  \href{garani@fi.infn.it}{garani@fi.infn.it}, \href{mailto:michele.redi@fi.infn.it}{michele.redi@fi.infn.it}, \href{andrea.tesi@fi.infn.it}{andrea.tesi@fi.infn.it}}}}
\newpage

\tableofcontents

\setcounter{footnote}{0}

%=========================================================================

\section{Introduction}

All the current evidences for the existence of dark matter (DM) rely on its gravitational interactions.
At late time the most convincing  observations are gravitationally collapsed structures ranging from the smallest 
known galaxies to the largest known galaxy clusters~\cite{Zyla:2020zbs} while CMB strikingly confirms this picture at early times \cite{Hinshaw:2012aka}. 
Lacking any non-gravitational evidence for DM,  truly dark sectors that perhaps interact only gravitationally are a plausible possibility.
To this end, pure gravitational production of dark sectors has gained much attention in recent times, and several models have been studied~\cite{Garny:2015sjg,Babichev:2016hir,Tang:2016vch,Garny:2017kha,Garny:2018grs,Ema:2018ucl,Bernal:2018qlk,Ema:2019yrd,Redi:2020ffc,Ahmed:2020fhc}. 

In the landscape of secluded dark sectors one possible hint to the nature of DM is the idea of accidental DM stability.
Stable particles are at the heart of the visible universe. While the electron is exactly stable, the proton is known to have an extremely long lifetime, $\tau>10^{34}$ yr. 
In the standard model (SM) proton stability is beautifully explained by accidental baryon number conservation of the SM lagrangian that is only violated by dimension 6 operators which are suppressed by the scale of new physics, perhaps the unification or the Planck scale. The same level of elegance is typically not shared by the dark sector where DM stability is often granted 
imposing ad hoc global symmetries, most notably R-parity in supersymmetry. 
Motivated by this observation, models with composite DM have been proposed in the literature~\cite{Nussinov:1985xr,Carlson:1992fn,Kribs:2009fy,Hietanen:2013fya,Hardy:2014mqa,Antipin:2015xia,Appelquist:2015yfa,Cline:2016nab,Lonsdale:2017mzg,Mitridate:2017oky,Carvunis:2020exc} (see also~\cite{Kribs:2016cew} for a review on the topic).

A large fraction of previous studies focused on dark sectors charged under the SM, leading to DM phenomenology in the same universality class of weakly interacting massive particles.
In the most compelling scenario DM is a baryon of the dark sector  with electro-weak charges \cite{Antipin:2015xia}, whose mass is expected to be around 100 TeV if thermally produced or lower  if asymmetric \cite{Bottaro:2021aal}. In this work we consider dark gauge sectors with fermions neutral under the SM.\footnote{For a discussion of secluded U(1) models we refer the reader to Refs.~\cite{Garny:2015sjg,Garny:2017kha}.}
This framework yields DM candidates in the form of dark hadrons which are characterized by a compositeness scale $\Lambda$ and by the absence of any renormalizable interactions with the SM. 
Concretely we will study SU($N$) gauge theories with $N_F$ light flavors, that we dub dark QCD (dQCD), see \cite{Dondi:2019olm,Morrison:2020yeg,Tsai:2020vpi} for related work.
The physics is described by the lagrangian,
\be\label{eq:definition}
\int d^4x \,\sqrt{-g} \,\bigg[ \mathscr{L}_{\rm SM}   -\frac{1}{4} G^a_{\mu\nu} G^{\mu\nu a} \, + \, \bar{\psi}_i \,(\slashed{D}-m_i) \,\psi_i  + \sum \frac{\mathcal{O}_{\rm SM}\mathcal{O}_{\rm dark} }{\Mpl^\#}\bigg]\,.
\ee
where ${\cal O}_{\rm SM}\, ({\cal O}_{\rm dark})$ are gauge invariant operators of the SM (dark) sector.

In the most minimal case with no fermions the lightest glueballs are accidental DM candidates \cite{Cline:2013zca,Boddy:2014yra,Soni:2016gzf,Forestell:2017wov,Acharya:2017szw,Jo:2020ggs}. In this context, gravitational production leads to viable models with dark sectors much colder than the SM ~\cite{Redi:2020ffc,Gross:2020zam}.

The addition of fermions gives rise to a host of new possibilities and novel phenomena. Accidentally stable DM candidates are in this case the lightest dark baryons and pions, where the latter
can be arbitrarily lighter than $\Lambda$ in principle. Baryon DM in dQCD was also studied in \cite{Dondi:2019olm} with somewhat different conclusions. 
Contrary to the pure glue scenario the leading interaction between the dark sector and the SM is through the Higgs portal,  $|H|^2\bar{\psi}\psi$. 
This dimension 5 operator even when suppressed by the Planck scale dominates DM production and controls pions stability. The latter turn out to be cosmologically stable if their mass is below the GeV scale. Gauge theories with fermions are approximately Weyl invariant except for the fermion mass terms and confinement scale. This implies that inflationary production is  suppressed if the Hubble scale during inflation is larger than $\Lambda$.  These sectors are instead populated through tree level gravitational interactions leading to sectors colder than the SM. It is also possible that heavy degrees of freedom provide  thermal contact with the SM or that the inflaton has a sizable width in the dark sector producing warmer sectors.

Depending on the dark sector initial temperature different scenarios emerge.
We will show how dark baryon (pions) can constitute heavy (light) DM, spanning many orders of magnitude in mass, can be realized in dQCD. 
Contrary to models with SM charges  the pions turn out to be excellent DM candidates beside baryons. This setup naturally endows DM candidates with intrinsic self-interactions at low energies, that leads to the exciting possibility of observable effects in structure formation. Thus we illustrate how completely secluded dark sectors could be phenomenologically tested through cosmological probes.  
In particular dark sectors that are initially in thermal equilibrium with the SM are generically excluded.

The paper is organized as follows. In Section~\ref{sec:dqcd} we review the basic features of confining SU($N$) gauge theories, and describe few properties of the lightest dark-baryons and -pions. In Section~\ref{sec:production} we describe in detail how dark sector particles could be produced in the early universe. We outline three such possibilities namely, gravitational production of CFTs, production through Planck suppressed operators, and inflationary production. In Section~\ref{sec:dqcd-hist} we follow the thermal history of the dark sector. We first consider a phase transition in the dark sector that results in confinement, and then proceed to the computation of the relic abundance for dark-baryons and -pions.  In Section~\ref{sec:pheno} we study the phenomenology of our setup and present the main results. Broadly, we find three viable regions of parameter space where confined dQCD would provide DM candidates. Finally, we draw our conclusions in Section~\ref{sec:conclusions}.   
In the appendix we present  general formulae for the production of a dark sector through the Higgs portal.

%%%%%%%%%%%%%%%%%%%%%%%%%%
%%%%%%%%%%%%%%%%%%%%%%%%%%
\section{Dark QCD}\label{sec:dqcd}

We are interested in asymptotically free non-abelian gauge theories with fermions lighter than the dynamical scale $\Lambda$.
This can be realized with classic SU($N$), SO($N$) and Sp($N$) gauge theories with vector-like fermions. 
For simplicity we focus on SU($N$) gauge theories with $N_F$ light flavors described by the  lagrangian (\ref{eq:definition}).
At energies much larger than the confinement scale $\Lambda$ this system is weakly coupled and approximately Weyl invariant. 
Assuming standard QCD dynamics this sector confines producing hadrons. The dynamics is such that the global chiral symmetry  SU($N_F$)$\times$ SU($N_F$)  
spontaneously breaks to the diagonal subgroup SU($N_F$) producing Nambu-Goldstone bosons, the pions. 
In this paper we assume that all the fundamental fermions $\psi_i$ are singlet under SM so they will have in general non-degenerate masses, $m_i \bar{\psi_i} \psi_i$,  preserving individual species number $\psi_i \to e^{i \alpha_i}\psi_i$. In practice we will consider degenerate quark masses for simplicity.
Among the hadrons we will focus on the pions that are the lightest states and baryons that are the natural DM candidates
being accidentally stable.

\paragraph{Pions:}~\\
The spontaneous breaking of chiral symmetry produces  $N_F^2-1$ light Nambu-Goldstone bosons in the adjoint of $SU(N_F)$. 
Their interactions are described as in QCD by the corresponding chiral effective lagrangian of the form
\be
\mathscr{L}_{\rm \pi} =\frac{f^2}4 \mathrm{Tr}(\partial_\mu U)^2 + b \mathrm{Tr}[M U + h.c.] + \mathrm{WZW},\quad\quad U=\exp[i \pi/f]\,{\rm and} \quad M_{ij}= m_i \delta_{ij}~.
\ee
where WZW is the Wess-Zumino-Witten topological term \cite{Witten:1983tw} and $\pi\equiv\pi^a T^a$, where $T^a$ are SU($N_F$) generators. Expanding the lagrangian above to fifth order in the pion field, for degenerate quark masses one finds
\be\label{eq:lagpi}
\begin{split}
\mathscr{L}_{\rm \pi}& =\frac 1 4 {\rm Tr}[\partial_\mu \pi \partial^\mu \pi]- \frac{M_\pi^2}4 {\rm Tr}[\pi \pi] + \frac{1}{48} \frac{M^2_\pi}{f^2} {\rm Tr}[\pi^4] - \frac{1}{24 \,f^2} {\rm Tr}[\pi \pi \partial_\mu \pi \partial^\mu \pi - \pi \partial_\mu \pi \pi \partial^\mu \pi] ~.  \nonumber \\
&+ \frac {N}{240 \pi^2 f^5} \epsilon^{\mu\nu\rho\sigma} {\rm Tr}[\pi \partial_\mu \pi \partial_\nu \pi \partial_\rho \pi \partial_\sigma \pi] +\dots
\end{split}
\ee
The interactions in the first line induce pion elastic scattering while the second line from the expansion of the WZW term is responsible for pion number changing processes \cite{Hochberg:2014kqa}.

\paragraph{Baryons:}~\\
The baryon spectrum is obtained by simply generalizing the eight-fold way of the strong interactions. The lightest multiplets are \cite{Antipin:2015xia}:
\begin{itemize}
\item for $N_F=1$, baryons are spin $N/2$ antisymmetric combinations of $N$ quarks.
\item For $N_F$ even, baryons are spin 0 particles in the symmetric representation of the flavor group.
\item For $N_F$ odd, baryons are spin 1/2 particles in the octet-like representation of flavor.
\end{itemize}
With our normalization, using QCD values, the lightest baryons are expected to have masses $M_B\sim 10 f$.

\subsection{Cosmological stability}
The dark sector is invariant under a global U(1) dark-baryon number. As a consequence the lightest dark baryon is accidentally stable. The lightest pion is also stable
as it is lightest state of the dark sector.  Allowing for higher-dimensional operators can in principle make dark baryon (pion) unstable over cosmological time scales.

In the absence of light right-handed neutrinos, we have two possibilities to break the above mentioned dark-baryon number. They involve SM operators $|H|^2$ and $LH$ as the dark sector is a singlet under the SM. For the lightest baryon $B$, there are two effective interactions that lead to decay, $B |H|^2$ ($B L H$) for even (odd) number of dark colors $N$.\footnote{For the special case $N_F=1$ baryons are higher spin, therefore the effective operators must contain extra derivatives leading to slower decays. For $N$ odd in the presence of right-handed neutrinos  we can also write $B \nu_R$, this however leads to suppressed decay within the standard see-saw mechanism.} The most constraining situation arises when $N=3$ as $BLH$ originates from a dimension 7 operator $\Psi^3 L H/\Mpl^3$. This results in baryon lifetimes that are compatible cosmological stability of DM,

\begin{equation}\label{eq:life_baryon}
\tau_B \sim  \frac {8\pi \Mpl^6}{M_B^7} \sim 10^{27}\,{\rm s} \left(\frac{4\times 10^8\rm GeV} {M_B}\right)^7\,.
\end{equation}

Dark pions can  decay through dimension 5 and 6 operators of the form,\footnote{Note that operators coupling to SM bilinear $f_i f_j$ are not allowed for dark sectors made of SM singlets. If right-handed neutrinos exist
the decay of pions are chirally suppressed by the mass ratio of light and heavy neutrinos, within the see-saw mechanism.}

\begin{equation}\label{eq:eff_op}
\frac 1 {\Lambda_5} \bar \Psi^i \gamma^5 \Psi^j |H|^2 + \frac 1 {\Lambda_6^2} \bar \Psi^i \gamma^\mu \gamma^5 \Psi^j \bar{f} \sigma^\mu f\,.
\end{equation}
These operators allow the lightest pion to decay to the SM, since they break individual species number. 
We focus on the Higgs portal operator that owing to its dimensionality produces the largest effects. Using $\langle 0| \bar \Psi \gamma^5 \Psi | \pi\rangle = c\, 4\pi f^2\,$
this generates the effective operator,
\begin{equation}
c   \frac{4\pi f^2}{\Lambda_5} |H^2| \pi~.
\end{equation}
For $M_\pi > M_H/2$ the dark pions can decay into on-shell Higgs. For $M_\pi < M_H/2$ the leading effect is due to the mixing with the Higgs boson, see \cite{Mitridate:2017oky}.
The most relevant decay are tree-level decay to SM fermions and 1-loop decay to photons when the pion is lighter than electrons. One finds,
\begin{equation}
 \Gamma_{\pi\to f\bar f}= N_c\frac {M_\pi}{16 \pi} y_f^2 \sin^2 \alpha \,, \qquad  \, \Gamma_{\pi \to \gamma\gamma}= \frac{\alpha^2}{256\pi^3}\frac{M_\pi^3}{v^2}c_\gamma \sin^2\alpha \,, 
\end{equation} 
where $\sin \alpha \approx c 4\pi  f^2 v/(\Lambda_5 M_H^2)$ and the coefficient $c_\gamma$ can be found for example in \cite{Djouadi:2005gi}. For $M_\pi < M_e$, neglecting confinement effects one finds $c_\gamma=121/9$. This leads to the lifetimes,
\begin{eqnarray}
\label{eq:fast}
\tau_\pi \big|_{M_\pi > 2 M_H} &\approx& \frac{M_\pi \Lambda_5^2}{ 2\pi c^2f^4} \approx 1 {\rm s} \,\,  \bigg(\frac{10\, \TeV}{\sqrt{c}f} \bigg)^4 \bigg(\frac{\Lambda_5}{\Mpl} \bigg)^2\bigg(\frac{M_\pi}{10\, \TeV}\bigg)\,,\\
\label{eq:intermediate}
\tau_\pi \big|_{ 2 M_e <M_\pi < 2 M_H} &\approx& \frac {20}{M_\pi} \frac {\Lambda_5^2 M_H^4}{c^2f^4 v^2}\frac {1}{y_f^2} \approx 2 \times 10^8\,{\rm s} \left( \frac{{\rm TeV}}{\sqrt{c} f}\right)^4 \left( \frac{{\Lambda_5}}{\Mpl}\right)^2\left( \frac{{M_b}}{M_f}\right)^2  \left( \frac{10\,{\rm GeV}}{M_\pi}\right) \,, \\
\label{eq:slow}
\tau_\pi \big|_{ M_\pi < 2 M_e} &\approx&  \frac 4 {M_\pi^3}\frac {\Lambda_5^2 M_H^4}{\alpha^2 c^2f^4} \approx 5 \times 10^{24}\,{\rm s} \left( \frac{{\rm TeV}}{\sqrt{c} f}\right)^4 \left( \frac{{\Lambda_5}}{\Mpl}\right)^2 \left( \frac{{\rm MeV}}{M_\pi}\right)^3\,.
\end{eqnarray}

These lifetimes determine the allowed region of parameters of dQCD.
If pions are DM conservatively their lifetime should be larger than $10^{26}$ s, see \cite{Arvanitaki:2008hq,Essig:2013goa} for a more detailed discussion. 
In practice since pion DM requires masses below GeV a weaker bound will apply.
If DM is made of baryons the pions can be a subdominant component of DM if they are cosmologically stable or they should decay before BBN, $\tau_\pi < 1 $ s.
Since in this case pions are only a fraction of DM a more detailed study is in principle required but we expect roughly similar constraints from decays.
Concerning baryons instead their lifetime does not lead to interesting constraints, at least in the regime where the DM abundance is reproduced. 

\begin{figure}[t]
\centering
\includegraphics[width=0.65\linewidth]{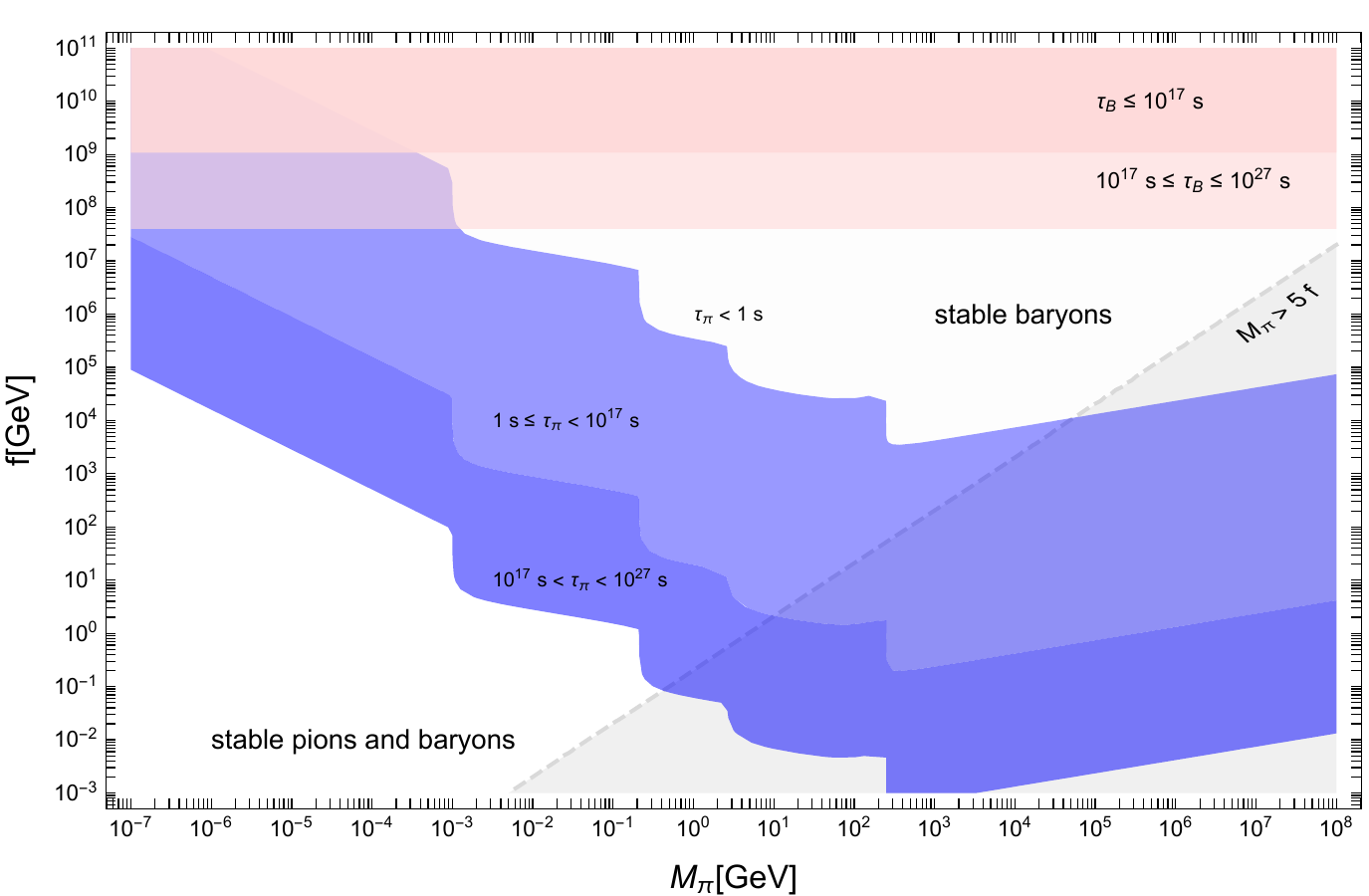}
\caption{\textit{2-D contours of the life time  of dark baryons (pink) and pions (blue) as a function of the pion decay constant $f$ and the pion mass $M_\pi$, with $\Lambda_5=\Mpl$ and $M_B=10\,f$. 
Blue region is excluded if the abundance of pions is comparable to the DM abundance. Massive pions decay through a dimension 5 operator, see text for details.}}
\label{fig:lifetime}
\end{figure}

The allowed region of parameter space is drawn in Fig. \ref{fig:lifetime}.

%%%%%%%%%%%%%%%%%%%%%%%%%%
%%%%%%%%%%%%%%%%%%%%%%%%%%

\section{Production mechanisms}\label{sec:production}
The dQCD sector we consider in this work is secluded from the SM. Since  the SM and the dark sector communicate only through gravity and  Planck suppressed higher dimensional operators, they are never in thermal equilibrium. It is therefore necessary to explain how the dark sector is populated in the early Universe. The possible mechanisms of production in our setup are sensitive to UV parameters, such as the Hubble scale during inflation, $H_I$, the reheating temperature $T_R$, or the inflaton couplings. Generally, dark sectors could be produced in the following ways:

\begin{itemize}
\item[1.] Production from the SM plasma: through exchange of graviton at tree level~\cite{Garny:2015sjg,Garny:2017kha,Ema:2018ucl} and at loop level~\cite{Tang:2016vch}, as well as through dimension 5 operators suppressed by the Planck scale via the Higgs portal, as we discuss in this paper.
\item[2.] Inflationary production: through quantum fluctuations in an expanding background \cite{Chung:1998zb}. This mechanism requires explicit breaking of Weyl invariance. In our model at high energies, this breaking is proportional to the fermion masses leading to very small effects as we will show below. 
\item[3.] Inflaton decays: the dark sector could be produced with an energy density $\propto \rho_D/\rho_{\rm SM}=\Gamma_D/\Gamma_{\rm SM}$, i.e. the production  is proportional to the inflaton branching ratio to the dark sector.
\item[4.] Renormalizable interactions: if heavy fermions charged under the SM exist then the system will be in thermal equilibrium with the SM 
at temperatures above the mass. 
\end{itemize}
In the rest of this section we discuss the first two production mechanisms and provide for them rather general results.

\subsection{Tree level production}

Any dark sector can be produced from the SM thermal bath as long as it has some feeble interactions~\cite{Hall:2009bx}. The only unavoidable production mechanism is through tree-level graviton exchange. In our scenario however higher-dimensional operators (suppressed by the Planck scale) can also give sizable production rates. We qualitatively discuss these possibilities below.

\paragraph{Gravitational production}~\\
The yield of gravitationally produced relativistic particles is given by \cite{Redi:2020ffc}
\begin{equation}
Y_{\rm D} = 6 \times 10^{-6} \, c_{D} \bigg(\frac{T_R}{\Mpl}\bigg)^3\,.
\label{eq:yield}
\end{equation}
Where $c_D$ is the central charge of the dark sector and $T_R$ the reheating temperature.\footnote{For a real scalar, Weyl fermion and massless gauge field the values of the central charges are respectively $c_0=4/3$, $c_{1/2}=4$ and $c_1=16$, see \cite{Osborn:1993cr}.}.
The presence of the central charge is due to the fact that at $T_R$ both the SM and dark sector are well approximated by relativistic CFTs, and given the tensor structure of the gravitational coupling, the production rate can be computed simply in terms of the two-point functions of the stress-energy tensor~\cite{Gubser:1997se}. Gravitational production gives rise to a dark sector that is under-populated compared to thermal equilibrium.
The typical energy of the dark quanta produced are however of order of the temperature of the visible sector so that the energy density is of order
$\rho_D\approx c_D T^4 T_R^3/\Mpl^3$. More precisely, solving the relevant Boltzmann equation, and neglecting small corrections from quantum statistics, one finds the following phase space distribution~\cite{Redi:2020ffc}
\begin{equation}
f_D(T,p)\approx \frac  {2\pi^4 g_*} {135} Y_D \frac {p \,  e^{-p/T}}T\,,
\label{eq:DMdistribution}
\end{equation}
which is only marginally different from a thermal Boltzmann distribution.

\paragraph{Planck suppressed operators}~\\
The abundance of particles that are produced from scatterings mediated by Planck suppressed effective operators can be comparable or even more important than graviton exchange depending on the dimensionality of the operator. However, contrary to gravitational production these contributions are model dependent, determined by the UV completion of the model. As a conservative assumption we will thus allow for higher dimensional operators suppressed by the Planck scale. The most phenomenologically relevant case arises when we consider the lowest dimensional operator in the SM, i.e. $\mathcal{O}_{\rm SM}= |H|^2$ (Higgs portal). Here we show the production rate for this case by considering a general $d$-dimensional operator in the dark sector $\mathcal{O}$ as follows
\begin{equation}
\frac {1} {\Lambda_{\rm UV}^{d-2}} |H|^2 {\cal O}\,,~~~~~~~~~~~~[{\cal O}]=d\,,
\end{equation}
where the effective scale  $\Lambda_{\rm UV}\sim \Mpl$. As reviewed in appendix~\ref{app:hgportal}, in the relativistic limit, scale invariance implies that the 2-point function of ${\cal O}$ is given by
\be
\langle \op(x)\op(0)\rangle =\frac{a_\op}{8\pi^4}\frac{1}{(x^2)^d}\,.
\ee
The annihilation cross section is then given by
\begin{equation}\label{eq:thermalxsec}
\langle \sigma v \rangle=\frac 1 {g_i^2} \frac{a_{\cal O}}{4 \pi}  \frac {T^{2d-6}}{\Lambda_{\rm UV}^{2d-4}}\,.
\end{equation}

The abundance of dark sector particles is obtained by solving the Boltzmann equation (see appendix~\ref{app:hgportal}), we get
\begin{equation}
Y_D=\int_0^{T_R} \frac {dT}T \frac {\langle \sigma v \rangle s}H Y_{\rm eq}^2 =a_{\cal O} \frac {135 \sqrt{5/2}}{4( 2 d-5) g_*^{3/2} \pi^8}\left( \frac  {T_R}{\Lambda_{\rm UV}}\right)^{2d-4}\frac {\Mpl}{T_R}\,.
\end{equation}
Assuming $\Lambda_{\rm UV}=\Mpl$ in the above, the dark sector operator of dimension three will result in the largest yield. This observation is relevant since such operators exist in the dQCD case considered in this work, the fermion bilinears $\mathcal{O}=\bar\Psi_i \Psi_j$ and $\mathcal{O}=\bar\Psi_i \gamma_5 \Psi_j$. Such $d=3$ operators are also responsible in making the dark-pion unstable. 

\paragraph{Inflaton scattering}~\\
The dark sector could also be produced through tree-level inflaton scatterings during reheating \cite{Ema:2018ucl,Mambrini:2021zpp}. This contribution is negligible in our case 
in light of the classical Weyl invariance of the action. As shown in appendix \ref{app:inf_scat} for a traceless CFT the production rate is identically zero. 
This implies that for elementary fermions and conformally coupled scalars the production is suppressed by the mass of the particle as found in explicit computations.
When the mass gap is created dynamically the suppression is even stronger because the dynamical scale should not be treated as a mass until the temperature drops below $\Lambda$.

\subsection{Inflationary production}
A different production mechanism relies on quantum fluctuations, usually called inflationary fluctuations, that can be relevant during (and after) inflation. We assume here $\Lambda< H_I$ so that the dark sector is deconfined during inflation. The basic principle here is the time-dependence of the metric background, which induces a change in the vacuum state interpreted as particle production (see for example \cite{Kolb:2020fwh} for a recent discussion and references).
Such an effect is very reduced if the theory enjoys an approximate Weyl symmetry, through which the time-dependence of the metric can be (classically) removed. This is the case of deconfined gauge fields and fermions, that are Weyl invariant in the massless limit. This implies that particle production must be proportional to the breaking of Weyl invariance, due to confinement at $H\sim \Lambda$,during reheating or radiation domination. For this reason, in our case we expect that the details of reheating will be important, but the overall contribution  to energy density is small as we now discuss. 
This discussion differs from the one in \cite{Gross:2020zam}.

Within the inflationary context, several works have studied the gravitational production of elementary particles\footnote{We focus the simplest possibility where the fermions do not couple to the inflaton. Such couplings may lead to new effects and different phenomenological predictions, see \cite{Adshead:2018oaa}.}.  It is interesting to summarize the results for an elementary fermion of mass $M$ \cite{Chung:2011ck,Ema:2019yrd,Herring:2020cah}. If the mass is small compared to Hubble at the end of inflation $H_e$, two contributions are possible depending on the value of Hubble at reheating $H_R$ \cite{Ema:2019yrd},
\begin{equation}
Y_D\approx \left\{ \begin{array}{ll} \displaystyle 5 \times 10^{-25}  \left( \frac M {3.3 \times 10^{11}\,{\rm GeV}}\right)^{\frac 5 2}\,\, & H_R \gg M\,, \\ \displaystyle 5 \times 10^{-22}  \left( \frac M {10^{9}\,{\rm GeV}}\right)\left(\frac {T_R}{10^{10}\,{\rm GeV}}\right)\,\, & H_e \gg M \gg H_R\,.\end{array}\right.
\end{equation}

While the precise abundance depends on the details of reheating, the crucial and robust result for the present discussion is that 
the production is dominated by co-moving momenta of order
\begin{equation}
\frac {k_c}a\approx M \approx H(a)\,.
\end{equation}
Since we assume $M< H_e$ this will happen during reheating or in radiation domination \cite{Chung:2011ck,Ema:2019yrd}:
\begin{equation}
\frac {k_c}{a_e}= M \left(\frac{H_e}M\right)^{\frac 2 3}\,~~~~~{\rm or}~~~~~~~~\frac {k_c}{a_e}= M \left(\frac{H_e}M\right)^{\frac 2 3}\left(\frac{M}{H_R}\right)^{\frac 1 6}\,.
\end{equation}
In other words, these modes are not produced by inflationary fluctuations but during reheating or radiation domination when the modes re-enter the horizon and they are  non-relativistic at production.

The discussion above makes it clear that for confining gauge theories the earlier estimates cannot be applied. 
Since by assumption $M < \Lambda$ one needs to take necessarily into account confinement. Similarly to finite temperature 
as long as $H/(2\pi)> \Lambda$ the physical degrees of freedom are gluons and quarks as the Hubble patch is smaller than the size of hadrons.
When $H/2\pi< \Lambda$ the physical degrees of freedom are the hadrons so that one should turn to a computation in terms of composites.  
Roughly in the equation above $M$ should then be replaced by the mass of composite states $g_* \Lambda$ if this is not too large. 
A more detailed computation could be done for the Goldstone bosons. However, we expect these effects to be rather small and subdominant compared to the contribution from vacuum misalignment which originates in a completely similar fashion to the case of axion-like particles. After confinement, the pion field will emerge with a random initial value, $\pi(x) \sim \Lambda$, in a given Hubble patch, which should be averaged upon. When Hubble drops below the mass of the pions $H_\pi\approx M_\pi$, they will begin to oscillate behaving as non-relativistic energy density. It is important to distinguish if oscillations occur during radiation domination or during reheating. In the latter case the entropy in the SM plasma increases, so we need to check whether $H_R \approx M_\pi$. The yield in the two branches are as follows

\be
M_\pi Y \approx \left\{ \begin{array}{ll}\displaystyle \frac{M_\pi^2 \Lambda^2}{s(T_\pi)} \approx \frac{0.4}{g_*^{1/4}}\frac{\sqrt{M_\pi} \Lambda^2}{\Mpl^{3/2}}<\frac{\Lambda^2}{\Mpl^2} T_R\,, & M_\pi < H_R \\
\displaystyle \frac{M_\pi^2 \Lambda^2}{s(T_R)}\frac{H_R^2}{H_\pi^2}\approx \frac{\Lambda^2}{\Mpl^2} T_R\,, & M_\pi > H_R  \end{array}\right.\,.
\ee
Note that in the second estimate above we have taken entropy dilution into account. We see that the first case gives a smaller final abundance, numerically we obtain
\begin{equation}
\Omega h^2 \sim \sqrt{\frac {M_\pi}{\rm GeV}}\left(\frac {\Lambda}{10^{10}\, {\rm GeV}}\right)^2\,, \quad M_\pi < H_R\,\,.
\end{equation}
This contribution will be negligible phenomenologically.

In the case of pure glue theories \cite{Redi:2020ffc} no light degrees of freedom analogous to pions exist.
In that case we expect the abundance from inflationary fluctuations to be given by the formulae above with  $M_\pi \to M_{\rm DG}\sim 5 \Lambda$.
Note that this estimate agrees parametrically with the energy released during the phase transition $\sim  \Lambda^4$.

\section{Thermal history}\label{sec:dqcd-hist}
The cosmological history of the dark sector begins with their production. As discussed in the preceding section, gravitational and inflationary mechanisms considered here lead to a non-thermal distribution of free quarks and gluons initially. Because of interactions in the dark sector as the universe expands the system is driven towards equilibrium. The issue of thermalization of gauge theories 
is general a complicated  problem see for example \cite{Arnold:2002zm,Kurkela:2014tea}. Here we will follow \cite{Redi:2020ffc} where a discussion of thermalization in a cosmological setting is presented in the context of pure glue gauge theories. Thermalization of the dark quark-gluon plasma is achieved through number changing processes such us $3\to 2$ interactions. For gravitational production the typical energy of the quanta is of order $T$
so that on dimensional grounds $\sigma_{2\to 3}\approx \alpha_{\rm eff}^3/T^2$ where $\alpha_{\rm eff}$ is the effective coupling that controls number changing processes. $\alpha_{\rm eff}$ can in principle be derived in terms of the perturbative gauge coupling and is thus related to the confinement scale. The number density in the dark sectors is roughly $n_D\sim Y_D g_* T^3$ so that the rate for number changing process $\Gamma \sim Y_D g_* \alpha_{\rm eff}^3 T$. Given that the Hubble rate during radiation domination scales as $H\sim T^2/\Mpl$ thermalization unavoidably occurs as long as the temperature is larger than the confinement scale.  We can  estimate the visible sector temperature where thermalization takes place as,
\begin{equation}
T_{*}\sim Y_D \Mpl \sqrt{g_*}  \alpha_{\rm eff}^3\,,
\label{Tstar}
\end{equation}
where $\alpha_{\rm eff}$ is the effective coupling that controls number changing processes, $\sigma_{2\to 3}\approx \alpha_{\rm eff}^3/T^2$. 
A posteriori we can check that the dark sector thermalizes in the deconfined regime when the DM abundance is reproduced. For this reason the only memory of the production 
mechanism is the ratio of temperature between visible and dark sector that we will take as input in the phenomenological analysis.

As the produced dark sector particles are relativistic, we use conservation of energy  and find that the dark sector temperature is given by
\begin{equation}
\xi^0\equiv \frac {T_D} T = \left(\frac{g_* \rho_D}{g_D \rho_{SM}}\right)^{\frac 1 4}\,,
\label{eq:TD}
\end{equation}
where $T$ ($T_D$) is the visible (dark) sector temperature. With $g_*$ ($g_D$) being the relativistic degrees of freedom in the visible (dark) sector.  This reasoning is valid for the production from the thermal plasma (through graviton exchange or higher dimensional operators) and from inflaton decay. Whereas, for the case of inflationary production this argument does not apply, as dark sector particles are produced while non-relativistic as discussed in the previous section.

As mentioned before, dQCD is conformal at high energies, thus quantities such as the central charge ($c_D$), degrees of freedom ($g_D$), and the coefficients of 2-point functions are completely determined by $N$ and the number of light flavors ($N_F$), as follows
\begin{equation}\label{eq:constants}
\begin{split}
&c_D^{QCD}=16 (N^2-1) + 8 N_F\,, ~~~~~~~ g_D= 2(N^2-1)+4 N_F N\,,\nonumber \\
&a_{\bar{\Psi}\Psi}=8 N N_F\,,\quad \quad~~~~~~~~a_{F^2/4}= (N^2-1)24~.
\end{split}
\end{equation}

Depending on the production mechanism considered the initial dark sector temperature $T_D$ has different parametric dependence on the reheating temperature $T_R$, and other variables in the theory. For example, if the two sectors interact in a renormalizable way thermal equilibrium is always reached at earlier times (or large temperatures) then, $\xi^0_{\rm ren}=1$. Whereas, production via tree-level graviton exchange results in~\cite{Redi:2020ffc}

\begin{equation}\label{eq:xi0gr}
\xi^0_{gr} \approx 0.3 \left(\frac{T_R}{\Mpl}\right)^{\frac 3 4}~.
\end{equation}
If effective operators with $d=5$ dominate the production, then the initial temperature ratio depends more weakly on the reheating temperature, 
\begin{equation}
\xi^0_{|H|^2}\approx 0.2 \left(\frac{a_{\bar{\Psi}\Psi}}{g_D} \frac{\Mpl T_R}{\Lambda_{\rm UV}^2}\right)^{\frac 1 4}\sim 0.1 \left(\frac{T_R}{\Lambda_{\rm UV}}\right)^{\frac 1 4}~.
\end{equation}\label{eq:xi0dim5}
Finally, if produced through inflaton decay $\xi^0$ is proportional to the inflaton branching in the dark sector,  
\begin{equation}
\xi^0_{\rm dec}= \left(\frac{g_* \Gamma_D}{g_D \Gamma_{SM}}\right)^{\frac 1 4}~.
\end{equation}
Let us note that the inflaton decay can generate a dark sector temperature larger than the SM one. 
As we will discuss below this possibility is rather strongly constrained.
With the initial dark sector temperature determined we now proceed to the discussion of confinement and phase transition in the dark sector.

\subsection{Dark phase transition}

Before we evaluate the abundance of DM candidates, we must first consider the nature of phase transition in our dQCD model. We assume that the sector has thermalized in the relativistic regime (unconfined phase). Similarly to ordinary QCD, as the dark sector temperature $T_D$ drops below some critical temperature the dark sector confines, resulting in color singlet states such as dark-baryons and dark-pions. 

Depending on $N$ and $N_F$ the phase transition could be first-order, or a cross over. A few comments are necessary at this point. Most studies often focus on the dynamics of pure-gluonic theory as they are relevant from a fundamental perspective, i.e. they capture the essential qualitative features of the phase transition, and they are numerically more convenient~\cite{Panero:2009tv}. In this case it is found that the phase transition is first order~\cite{Lucini:2012gg} (and references within) from lattice calculations. Essentially, transition occurs without super-cooling and in equilibrium, resulting in a small increase of entropy~\cite{Brambilla:2014jmp}.

Inclusion of light fermionic degrees of freedom can change the above picture qualitatively~\cite{AliKhan:2001ek,Aoki:2005vt}. In this case phase transition can be first order or cross-over depending on $N_F$. It is found that the transition is weakly first order for $3\le N_F\lesssim 4 N$ for $N>3$~\cite{Brambilla:2014jmp}, which is expected to be adiabatic with the critical temperature $T_c \simeq \mathcal{O}(1)\,f $~\cite{Borsanyi:2012ve}.

As the phase transition completes the system reorganizes in color neutral states. We call $T_\Lambda$ the SM temperature when this happens. In this confined phase baryons and pions are the physical degrees of freedom. As pions are quite light ($M_\pi \lesssim 5\,f$), they are relativistic at production in the dark plasma. Consequently their interactions grow with energy, such that they are sufficiently fast and equilibrate. Baryons on the other hand are heavy with $M_B \sim 10\,f$. Nevertheless they also thermalize with dark thermal bath, but their abundance is suppressed at the phase transition as the temperature is smaller than $M_B$.

If the phase transition is a cross-over or sufficiently adiabatic we can use entropy conservation to determine the temperature after the phase transition.\footnote{In~\cite{Garcia:2015loa} the temperature after the phase transition  was determined through energy conservation assuming that the phase transition of pure glue theories is explosive, increasing the entropy. We believe a transition in quasi-equilibrium is more plausible for QCD-like theory. Nevertheless  the two conditions lead to a similar temperature after the phase transition.}
One finds that the ratio of temperatures $\xi\equiv T_D/T|_{T_\Lambda}$ right after the phase transition is given by, 
\begin{equation}
\frac {\xi}{\xi_0}\approx  \left(\frac {2 (N^2-1)+4 N N_F}{N_F^2-1} \right)^{\frac 1 3}\,.
\label{eq:xiprime}
\end{equation}

Let us discuss the case when the interactions are not sufficiently strong to thermalize in the relativistic regime, i.e. $T_*< \Lambda$ in eq. (\ref{Tstar}).
If the pions are relativistic right after the phase transition they will thermalize. The leading number changing process 
is due to the Wess-Zumino-Witten term $N/(4\pi)^2 \partial^4\pi^5/f^5$ that induces $2\to 3$ processes. 
This leads to number changing processes with cross-sections $\sigma_{2\to 3} \sim N^2 T^8/f^{10}$ which are  unsuppressed at the phase transition 
leading to rates faster than Hubble. In this case we can estimate the  temperature using conservation of energy as in eq. (\ref{eq:TD}).

\subsection{Dark sector temperature after the phase transition}

On the completion of the phase transition the dark baryons and pions thermalizes with a ratio of temperature $\xi=T_D/T$ if rates are sufficiently fast. In this phase we take $T_D$ to be the temperature of the dark pion gas, which will be relativistic for (much) longer than baryons. In principle the dark sector can have a temperature evolution different for each (relativistic/non-relativistic) species. However, as baryons and pions interact with each other we expect their temperature to be the same, thanks to kinetic equilibrium. In principle, however, they might be different and can be defined as \cite{Yang:2019bvg,Mondino:2020lsc}
\be
T_i \equiv\frac{P_i(T_i)}{n_i(T_i)}=\frac{g_i}{n_i(T_i)}\int \frac{d^3p_i}{(2\pi)^3}\frac{p_i^2}{3E_i} f_i(T_i)\,.
\ee
where $P_i$ is the pressure of the $i$-th species.  By integrating the Boltzmann equation for the $i$-th species, with a weight $p_i^2/(3E_i)$, we get
\be
\frac{n_i}{T_i} \big(\dot T_i + \delta_i H T_i\big)= - \big(\dot n_i + 3 H n_i\big) + \frac{g_i}{T_i}\int d\pi_i C[f_i\cdot \frac{p^2_i}{3E_i}]\,\,.
\ee
With
\be \label{eq:delta}
\delta_i\equiv 1 + \frac{g_i}{T_i n_i }\int \frac{d^3p_i}{(2\pi)^3}\frac{ p_i^2}{3E_i} \frac{m_i^2}{E_i^2} f_i(T_i) = \left\{ \begin{array}{cc} 1 & T_i\gg m_i \\ 2 & T_i\ll m_i \, .\end{array}\right.
\ee
We consider the case where kinetic equilibrium is maintained in the dark sector, this enforces all temperatures to follow the one of dark pions, $T_i=T_D$. It is then useful to define the total number of dark sector particle, $n\equiv \sum_i n_i$, to recast the set of Boltzmann equations into a single one
\be
n \frac{\dot T_D}{T_D} + n H T_D + \sum_{i} n_i (\delta_i-1) H T_D \approx  - (\dot n + 3 H n)\,.
\ee
Given that pions are more abundant that baryons we can simply follow the evolution of the number density of pions. Moreover the total number of dark sector particles is approximately conserved during the freeze-out of baryons (we neglect possible cannibalistic effects when pions are non-relativistic), so that the evolution of the dark sector temperature $T_D$ only depends on pions being relativistic or non-relativistic. Therefore in terms of the visible temperature we expect the following behavior
\be\label{eq:temperature}
T_D(T)=\left\{ \begin{array}{cc} \displaystyle \left(\frac{g_*^s(T)}{g_*^s(T_\Lambda)}\right)^{\frac13} \xi\, T, &  ~~~~~~~T_D > M_\pi  \quad \mathrm{equivalent\ to\ } T_\Lambda \geq T>M_{\pi}/\xi 
\\  \displaystyle  \left(\frac{g_*^s(T)}{g_*^s(T_\Lambda)}\right)^{\frac23} \xi^2 \frac{T^2}{M_\pi}, & T_D < M_\pi \quad \mathrm{equivalent\ to\ } T<M_{\pi}/\xi \end{array}\right.\,,
\ee
where we have included the effect of the decoupling of SM species.

\subsection{DM abundance}\label{sec:abundance}

In what follows we assume that the system has thermalized  and $\xi$ is the ratio of temperatures of the dark sector and SM  after the phase transition.

\paragraph{Pion abundance:}
The numerical abundance of dark pions is mainly set by the phase transition.
If the phase transition takes place in approximate thermal equilibrium we can estimate the abundance of pions at the onset of the confined phase as
\begin{equation}\label{energy-pions}
n_\pi = (N_F^2-1) \frac{\zeta(3)}{\pi^2} T_{D*}^3\,, ~~~~~~~\rho_\pi = (N_F^2-1) \frac {\pi^2}{30} T_{D*}^4\,.
\end{equation}
The yield of pions (defined with respect to the SM entropy), at the end of the phase transition, is then given by
\begin{equation}\label{yield-pions}
Y_{\pi}= (N_F^2-1) \frac{45 \zeta(3)}{2\pi^4 g_*} \xi^3\,.
\end{equation}
We notice that, since the dark sector only consists of baryons and pions, and that the baryons are heavier than $T_{D*}$, the baryon abundance is already suppressed at the onset. The above yield (and energy density) of pions is mostly unaffected by the subsequent evolution. The reason for this is that pions will be relativistic during the freeze-out of baryons. There might be an exception if pions are subject to number changing processes while non-relativistic during and/after baryon freeze-out, however this 'cannibalistic' phase would only deplete the above yield by a logarithmic correction \cite{Carlson:1992fn}. If the pions are cosmologically stable eq.~\eqref{yield-pions} can be used to compute the DM abundance and the corresponding DM mass
\begin{equation}\label{eq:abpi}
\frac{\Omega h^2}{0.12}\approx 0.6 (N_F^2-1)\frac {M_\pi}{0.1 \GeV}\left(\frac{\xi}{10^{-2}}\right)^3 \left(\frac{106.75}{g_*}\right)\,\quad \to \quad M^{\rm DM}_\pi \approx \frac{0.14\, \keV}{N_F^2 \xi^3} \left(\frac{g_*}{106.75}\right)\,.
\end{equation}
The value of DM mass is compatible with cosmological stability if it does not exceed $O(100\, \GeV)$. Lighter DM masses are achieved in theories with large number of flavors, as $M_\pi^{\rm DM}\sim 1/N_F$. Which takes into account the dependence on $N_F$ in $\xi$ and $\xi^0$.

In realizations where $\xi^0 \ll 10^{-3}$ the prediction for DM mass is incompatible with stability as estimated in eq.~\eqref{eq:fast}. In this case the pions decay to the SM injecting entropy into the SM plasma diluting the abundance of baryons. We estimate this by computing the release of energy at a temperature $T_\Gamma\approx 1.9 (\Mpl \Gamma_\pi)^{2/3}/(g_* M_\pi Y_\pi)^{1/3}$ when $H\sim \Gamma_\pi$ (assuming matter domination, valid for $T_\Gamma < (4/3) M_\pi Y_\pi$), we derive the new entropy density of the SM and the corresponding dilution factor  $\eta$ as
\be\label{eq:dilution}
\eta=\frac{s}{s_{\Gamma}}\approx \mathrm{min}\bigg[1,\,\, 0.8 \frac{(T_\Gamma/M_\pi)^{3/4}}{Y_\pi^{3/4}}\bigg]\approx \mathrm{min}\bigg[1,\,\,\frac{1.3}{g_*^{1/4}} \bigg(\frac{\Mpl^2 \Gamma_\pi^2}{M_\pi^4}\bigg)^{1/4} \frac{1}{Y_\pi}\,\bigg].
\ee
Asymptotic yields of stable particles have to be multiplied by $\eta$. Numerically we implement the full calculation, and under the assumption of matter domination at the time of decay \cite{Cirelli:2018iax}. In order for the scenario to be viable we require that the reheating temperature after the entropy release from the decay of the pions to be $T_{R,\pi} \approx\sqrt{\Mpl \Gamma_\pi} \gtrsim 10\ \mathrm{MeV}$.

\paragraph{Baryon abundance:}
The relic abundance of baryon is set by annihilation into multi-pion final states and the corresponding Boltzmann equation reads,
\begin{equation}
\frac {d n_B}{dt}+ 3 H n_B=- \sum_n \langle \sigma v \rangle_n \left[n_B^2 - \left(\frac {n_\pi}{n_\pi^{\rm eq}(T_D)}\right)^n (n_B^{\rm eq}(T_D))^2\right]\,.
\end{equation}
Where we have allowed for annihilation to $n$ pions that are expected to dominate compared to two-pion processes~\cite{Amsler:1997up}. The equilibrium number density are computed in terms of the dark sector temperature given by eq.~\eqref{eq:temperature}.
In the relativistic regime  whether the pions are in equilibrium or not they will have thermal distribution at temperature $T_D=\xi T$ due to the initial 
conditions. This allows us to rewrite the equation above in terms of the total annihilation cross-section.
With the standard manipulation the equation above can be cast in the following form
\begin{equation}
\frac {dY_B}{dT_D}= \frac{\langle \sigma v \rangle s(T)}{H(T) T_D}\big[Y_B ^2- (Y_B^{\rm eq}(T_D))^2\big]\,,
\end{equation}
with $T_D$ being the dark sector temperature. 
Due to the linear relation between $T$ and $T_D$ this is just the standard 
freeze-out equation with an effective cross-section $\langle\sigma v\rangle/\xi^0$. It follows that the freeze-out takes place as usual when 
$T_d\sim M_B/25$ and the abundance is just re-scaled by $\xi$. 
\footnote{
The freeze-out condition on $T_D$ is still determined by the condition $\langle \sigma v\rangle n^{\rm eq}_B(T_D)\approx H(T)$, which gives a condition 
$ M_B/T_D\big|_{f.o.}\approx \log(M_B \Mpl \langle \sigma v \rangle/(5.2\sqrt{g_*})) + 2\log\xi$.
}
When the baryons are in thermal contact with the SM the abundance 
is obtained for masses around 100 TeV corresponding to annihilation cross-section $\sigma v \sim 25 /M_B^2$ .
In this case the abundance is given by,
\begin{equation}
\frac {\Omega_B h^2}{0.12} \approx \xi  \bigg(\frac{g_*+g_D \xi^4}{106.75}\bigg)^{1/2} \left(\frac {M_B}{100\, \rm TeV}\right)^2 \,\quad \to\quad M_B^{\rm DM} \approx \frac{100\, {\rm TeV}}{\sqrt{\xi}}\bigg(\frac{106.75}{g_*+g_D \xi^4}\bigg)^{\frac14}\,.
\label{eq:abB}
\end{equation}
Note that this mechanism allows us to easily avoid the unitarity bound of standard freeze-out. 
In the case of pion DM the baryon abundance is a negligible contribution to the energy density. The above formula does not take into account the possible dilution coming from the late decays of unstable pions (see eq.~\eqref{eq:fast}). In this case the final abundance has to be modified by a factor equal to eq.~\eqref{eq:dilution}. If pions inject sufficiently large entropy, the DM baryon mass estimate gets modified to
\be\label{eq:baryonDMmass}
M_B^{\rm DM} \approx \frac{100\, {\rm TeV}}{\sqrt{\xi}}\bigg(\frac{106.75}{g_*+g_D \xi^4}\bigg)^{\frac14}\mathrm{max}\bigg[ 1,\ \ 10 N_F \xi \bigg(\frac{g_*+g_D \xi^4}{106.75}\bigg)^{\frac18} \left(\frac {M_{\pi}}{10^4\,{\rm GeV}}\right)^{3/8}\bigg]\,,
\ee
where we used eq. (\ref{eq:fast}) for  pion decay through the Higgs portal. For the model under consideration, we notice that when dilution is important the dependence on $\xi$ drops out when $\xi\to \infty$. The abundance scales as
\begin{equation}
\frac {\Omega_B h^2}{0.12} \sim \frac {g_*^{1/4}}{g_D^{1/2}} \frac {\sqrt{\Mpl \Gamma_\pi}}{M_\pi}\left(\frac {M_B}{100\, \rm TeV}\right)^2\,.
\end{equation}
This corresponds to the limiting case where initially the SM sector is not populated and it entirely originates from the reheating of the plasma upon pion decay.

The above estimates need to be revised if the baryon freeze-out happens when the pions are also non-relativistic. In order for this regime to be relevant the pions have to be $M_\pi \gtrsim T_D|_{f.o}$, and we have two possible cases to consider depending on whether pion number changing processes are fast or not. 

\medskip

\textbf{Non-relativistic pions decoupled:} Assuming that pion number changing processes are irrelevant $n_\pi \propto 1/a^3$. Since  $v^2\propto 1/a^2$ the
effective temperature drops as $T_D \sim 1/a^2$. This implies that the second term in the Boltzmann equation is enhanced by $e^{n M_\pi/T_D}$.
Note that in this case the Boltzmann equation cannot be cast in terms of total annihilation cross-section and processes with the largest
number of pions are favored. By taking inspiration from nuclear physics data \cite{Amsler:1997up} we argue that the dominant channel is the one dominated by the largest number of pions allowed kinematically, $Q_{n,\rm max}=2 M_B - n M_\pi \to  0$. The Boltzmann equation then reduces simply to
\begin{equation}
\dot n_B + 3 H n_B\approx -  \langle \sigma v \rangle_{n, \rm max} \left[n_B^2 - \left(\frac {n_\pi}{n_\pi^{\rm eq}(T_D)}\right)^n (n_B^{\rm eq}(T_D))^2\right]\,.
\end{equation}
The loss of equilibrium is then set by the condition $\langle \sigma v \rangle_{n, \rm max} \left(n_\pi/n_\pi^{\rm eq}\right)^{n/2} n_B^{\rm eq} \approx H$, which, taking into account that $T=\sqrt{T_D M_\pi}/\xi$, gives a freeze-out temperature 
$Q_{n,\rm max}/T_D|_{f.o.}\approx 2\log(\langle \sigma v \rangle_{n, \rm max}  M_\pi \Mpl \xi^2)$. By solving the differential equation we get 
\be
\frac{\Omega_B^{(n,\rm max)}}{\Omega_B} \approx 0.1 \frac{M_B}{\sqrt{m_\pi Q_{n,\rm max}}}\frac{\langle \sigma v \rangle}{\langle \sigma v \rangle_{n,\rm max}}\,.
\ee
This shows that we do not expect a large deviation from the previous case, since the largest deviation appears for $M_\pi \gtrsim M_B/20$ and $Q_{n,\rm max}\approx M_\pi$, which does not lead to a large effect.\\

\textbf{Non-relativistic pions in equilibrium:} If the pion number changing processes are fast, then dark pions undergo `cannibalism' that will make $n_\pi$ track $n_\pi^{\rm eq}$ with a new scaling of the temperature $T_D(T)$ fixed by conservation of entropy. Since we are studying the freeze-out of baryons, we do not consider pion-number changing processes involving baryons, that will eventually decouple, but we focus on pion self-interactions induced by the WZW in the chiral lagrangian, that allows for $\pi\pi\pi\to \pi\pi$. If self-interactions decouple later than baryon annihilations, this corresponds to a `cannibal phase' \cite{Pappadopulo:2016pkp,Farina:2016llk}. We acknowledge the possible presence of this effect, but neglect the logarithmic corrections to baryon and pion abundance induced by cannibalism. Phenomenologically in the regions where $M_\pi\approx M_B$, pions are non-relativistic during baryon freeze-out, this will not play a major role in the following discussion.

\section{Phenomenology}\label{sec:pheno}

The phenomenology of our scenarios is essentially determined by  3 parameters,
\begin{equation}
f\,, ~~~M_\pi\,, ~~~ \xi\,.
\end{equation}
The dark pion decay constant $f$ is also roughly the temperature of the deconfinement phase transition 
while the mass of the baryon is $M_B \sim  10 f $. The relic abundance formulae for both baryons and pions (when they are stable) impose a relation between these quantities so that $\xi$ can be eliminated in terms of the other parameters. It is then useful to discuss these models in the plane $(M_\pi,\,f)$. We will discuss the phenomenology of our model in terms of these two parameters, determining $\xi$ from the DM abundance constraint.

\begin{figure}[t]
\centering
\includegraphics[width=0.85\linewidth]{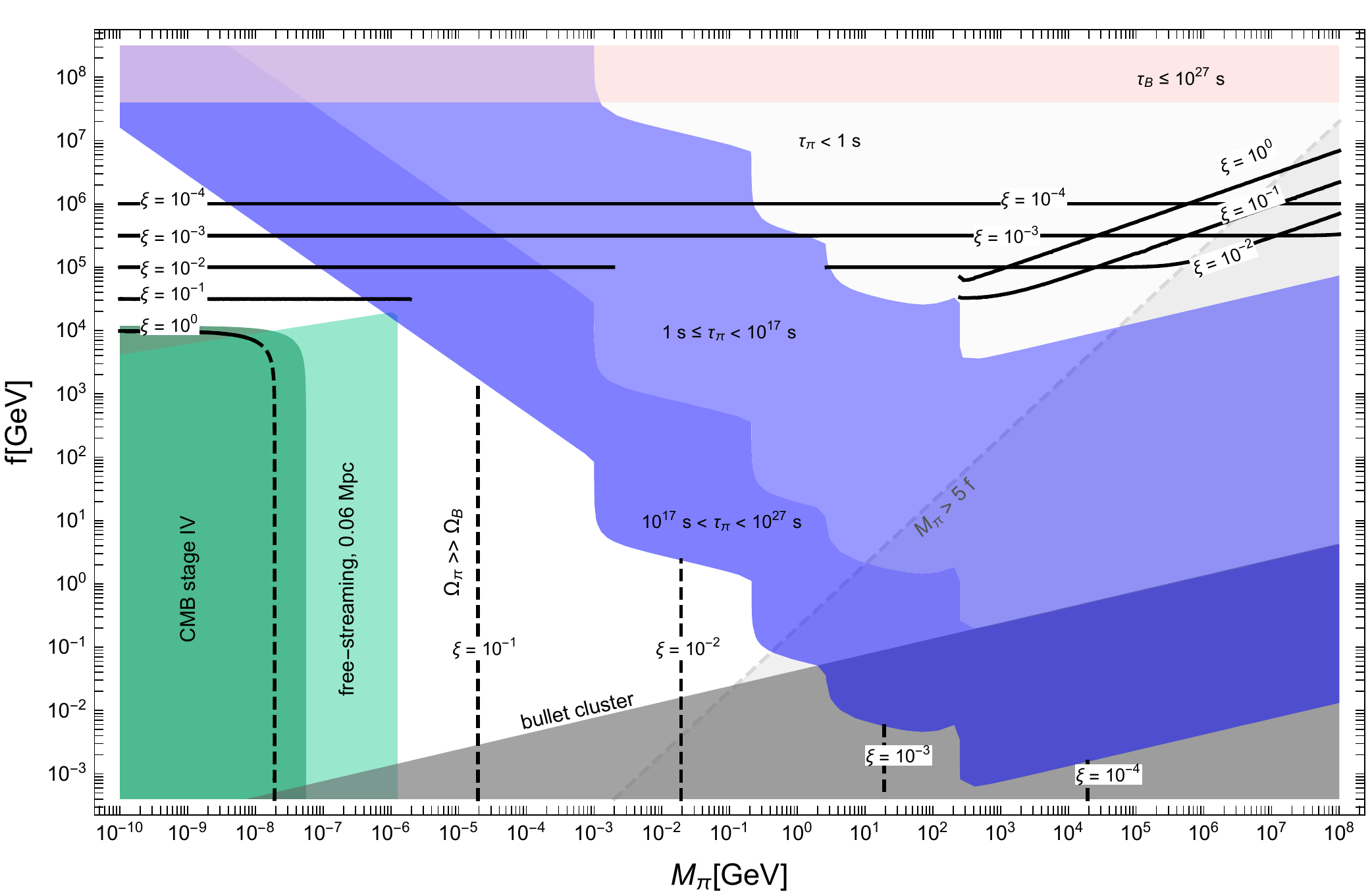}
\caption{\label{fig:abundance}\it  Parameter space of (secluded) dark QCD as a function of $M_\pi$ and $f$, with baryon mass fixed to $M_B=10 f$, $N_F=3$ and $\Lambda_5=\Mpl=2.4\times 10^{18}\,{\rm GeV}$. On the black solid isolines DM abundance is reproduced by dark baryons, while on the black dashed isolines by dark pions. Regions of stability are as in figure \ref{fig:lifetime}.}
\end{figure}

The results of the the phenomenological study are summarized in figure ~\ref{fig:abundance}.

In the rest of this section we discuss in detail the phenomenological implications of our model, and at the end we characterize the possible phases of our scenario.

\subsection{DM self interactions}

The elastic self scattering cross-section of dark baryons is expected to be at least geometrical and possibly enhanced by light di-baryon intermediate states. Re-scaling the QCD value one finds
\begin{equation}
\sigma^B_{\rm el}\approx \frac{4 \pi}{M_B^2}.
\end{equation}
The abundance of baryons reproduces the correct DM abundance only for extremely heavy baryons, $M_B \gtrsim 100\, {\rm TeV}$ (see Eq.~\eqref{eq:abB}), and this does not lead to any experimentally interesting constraints.  Of more interest are pions self-interactions, that could show up both in the Bullet Cluster and Ly-$\alpha$ constraints as we will discuss in the following. The leading interaction can be computed from the expansion of  the chiral lagrangian (\ref{eq:lagpi}), see~\cite{Hochberg:2014kqa}. The elastic cross-section at low energy is just,
\begin{equation}
\sigma_{\rm el}^{\pi}\simeq \frac{1}{64\,\pi} \frac{ (3 N^4_F -2 N^2_F + 6)}{N^2_F(N^2_F -1)} \frac{M^2_\pi}{f^4}\quad \stackrel{N_F \rightarrow 3}{=}\quad \frac{77}{1536\,\pi} \frac{M^2_\pi}{f^4}\,.
\end{equation}
We show regions (shaded in gray in figure \ref{fig:abundance}) that are excluded by the limit on DM self-scattering cross section ($\sigma^\pi_{\rm el}/M_\pi < {\rm cm}^2/{\rm g}$) from the Bullet cluster~\cite{Spergel:1999mh}.  
\subsection{Dark radiation}

If the quark masses are vanishing, $M_\pi\sim 0$, DM is made of baryons while pions contribute as dark radiation.
Assuming $T_\Lambda > 100$ GeV, the contribution to the relativistic number of degrees of freedom at the CMB epoch is
\begin{equation}
\Delta N_{\rm eff}\big|_{\rm CMB}= \frac 4 7\left(\frac {g_\nu}{g_{\rm eff}}\right)^{4/3} (N_F^2-1) \xi^4= 0.027(N_F^2-1) \xi^4\,,
\end{equation}
where $g_\nu=43/4$ is the number of degrees of freedom at neutrino decoupling. For $N_F=3$ and $\xi^0=1$ this leads to a contribution close to the experimental bound, which we take $\Delta N_{\rm eff}\lesssim 0.25$ at 95\% confidence level \cite{Fields:2019pfx}. This translate into a bound on the ratio of temperature $\xi \lesssim 1$. This bound is expected to significantly improve in the coming years to reach a 1$\sigma$ exclusion bound of $\Delta N_{\rm eff}< 0.027$ during the CMB stage IV \cite{Abazajian:2016yjj}. These regions are depicted in dark green in figure \ref{fig:abundance}.

\subsection{Structure formation}

Two of the key parameters that control structure formation are the free streaming and collisional damping scales \cite{Irsic:2017ixq}.  We discuss this for the case of pion DM, since baryons are usually sufficiently heavier than the mass scales probed by structure formation.
Here we summarize the relevant equation for free-streaming and collisional damping scales. Assuming that DM becomes non-relativistic at a time $a(t_{\rm NR})=a_{\rm NR}$ before matter-radiation equality, $a_{\rm eq}=a(t_{\rm eq})$,  we have the following expressions
\begin{eqnarray}
\lambda_{\rm FS} & = & \int_{t_{\rm dec}}^{t_{\rm eq}} \frac{dt}{a(t)} \langle v(t)\rangle  \,,\\
\lambda^2_{\rm coll.} & = & \int_0^{t_{\rm dec}}  \frac{dt}{a(t)^2} \frac{\langle v(t)\rangle^2}{n(t) \langle \sigma_{\rm el} v\rangle} \,.
\end{eqnarray}
In order to perform these integrals it is often convenient to change variables $dt=da/(a H(a))$, taking into account that deep in radiation domination $H(a)\approx H_0 (a_{\rm eq}/a)^2 (1/a_{\rm eq})^{3/2}$. We can derive analytical expression depending on whether decoupling occurs while DM is relativistic or non-relativistic. We identify the decoupling time through the relation $n(t)\langle \sigma_{\rm el} v \rangle \approx H(t_{\rm dec})$ as in \cite{Chu:2015ipa}, \footnote{\label{footnote:bullet}Requiring that the self-interactions decouple at matter-radiation equality gives an upper bound on $\sigma_{\rm el}/M_\pi$,
\be
\frac{\sigma_{\rm el}}{M_\pi} \bigg|_{\rm eq}\lesssim \frac{a_{\rm eq}}{a_{\rm NR}}  \frac{a_{\rm eq}^{3/2}}{3 H_0 \Mpl^2 \Omega_{\rm DM}} \approx 10^{-3}\, \frac{a_{\rm eq}}{a_{\rm NR}}\,  \frac{\mathrm{cm^2}}{\rm g} \approx 10^{-3}\, \frac{M_\pi/T_{\rm eq}}{\xi}\,  \frac{\mathrm{cm^2}}{\rm g} \,.
\ee
This bound has to be compared with the one arising from the Bullet Cluster and it is usually subdominant if DM becomes non relativistic early on. 
}
and we parameterize the DM velocity as $\langle v(t)\rangle=\theta(a_{\rm NR}-a) + (a_{\rm NR}/a) \theta(a-a_{\rm NR})$, which takes into account the non-relativistic behavior with a step function, see also \cite{Egana-Ugrinovic:2021gnu}. The scale factor when DM becomes non-relativistic is given by
\begin{equation}
a_{\rm NR}\approx\xi \frac {3 T_0}{M_\pi} \bigg(\frac{g^{s}_{*}(T_0)}{g^{s}_{*}(T_\Lambda)}\bigg)^{\frac13}\,.
\end{equation}
The above integrals are dominated by the time when DM becomes non-relativistic, and they scale with the above parameter, $a_{\rm NR}$. Indeed we can compute the above formulae in two different regimes for the DM velocity and find that the maximum scale is always proportional to $a_{\rm NR}$, with or without interactions. Depending on whether the decoupling occurs when DM is relativistic or not, one finds
{\small
\begin{eqnarray}
\lambda_{\rm FS} & =&  \frac{a_{\rm NR}}{H_0 \sqrt{a_{\rm eq}}} \bigg[ 1 -\frac{a_{\rm dec}}{a_{\rm NR}}+ \log \big(\frac{a_{\rm eq}}{a_{\rm NR}}\big)\bigg] \theta(a_{\rm NR}-a_{\rm dec})+\frac{a_{\rm NR}}{H_0 \sqrt{a_{\rm eq}}}  \log \big(\frac{a_{\rm eq}}{a_{\rm dec}}\big)\theta(a_{\rm dec}-a_{\rm NR}) \,,\\ 
\lambda_{\rm coll.} & =&  \frac{a_{\rm dec}}{ \sqrt{3}H_0 \sqrt{ a_{\rm eq}}} \theta(a_{\rm NR}-a_{\rm dec}) + \frac{a_{\rm NR}}{\sqrt{2}H_0 \sqrt{a_{\rm eq}}} \bigg( 1 - \frac{1}{3}\frac{a_{\rm NR}^2}{a_{\rm dec}^2} \bigg)^{\frac12}\theta(a_{\rm dec}-a_{\rm NR}) \,.
\end{eqnarray}}

We see that the largest scale is always of the order of $\lambda_{\rm max} \approx a_{\rm NR}/(H_0 \sqrt{a_{\rm eq}})$, while $a_{\rm dec}$, and hence the self-interactions, only gives a subleading correction. This is similar to the case analyzed in \cite{Egana-Ugrinovic:2021gnu}, where self-interactions give negligible effects unless they are large enough to be in tension with the Bullet Cluster constraint (see our footnote \ref{footnote:bullet}).

Therefore with this observation, and for the level of our discussion, it is enough to compute the free-streaming length. By retaining the relevant parametric dependencies we find the following approximate expression
\be
\lambda_{\rm FS}  
\approx 0.3\, \mathrm{Mpc} \frac{\mathrm{KeV}}{M_\pi/\xi} \bigg(\frac{106.75}{g^{s}_{*}(T_\Lambda)}\bigg)^{\frac13}\,.
\ee
We notice that the SM temperature when pions become non-relativistic is  approximately $T_{\rm NR}\approx \xi^{-1}(M_\pi/3)(g_*^s(T_\Lambda)/g_*^s(T_{\rm NR}))^{\frac13}$, which can be much smaller than $\mathrm{MeV}$, so that it makes sense to consider $g_*^s(T_{\rm NR})=g_*^s(T_0)$. See also \cite{DEramo:2020gpr} for an analytic estimate of the above quantity. An upper bound on $\lambda_{\rm FS}$ arises from the study of the Ly-$\alpha$ forest \cite{Irsic:2017ixq}, assuming 100\% of DM. By using eq.~\eqref{eq:abpi} we can therefore impose the relic abundance constraint and we get the following expression for the free-streaming length
\be
\lambda_{\rm FS}\big|_{\Omega_{\rm DM}} \approx 20 \, \mathrm{Mpc}\, \xi^4 \,\bigg(\frac{N_F^2-1}{8}\bigg)  \bigg(\frac{106.75}{g^{s}_{*}(T_\Lambda)}\bigg)^{\frac13}\,.
\ee
We see that the effect decouples fast when $\xi\ll 1$. Upon imposing the constraint from the relic abundance we see that $T_{\rm NR}|_{\Omega_{\rm DM}} \approx  60\, \mathrm{eV}/(N_F^2 \xi^4) \times (g_*(T_\Lambda)/106.75)\times (g_*^s(T_\Lambda)/g_*^s(T_{\rm NR}))^{\frac13}$, which for the region relevant for Ly$-\alpha$ constraints is certainly below the $e^+e^-$ decoupling temperature, as expected. Phenomenologically we impose the constraint $\lambda_{\rm FS}^2+\lambda_{\rm coll.}^2\lesssim (0.06\, \mathrm{Mpc})^2$~\cite{Chu:2015ipa}. These regions are shown in light green in figure~\ref{fig:abundance}.
\subsection{Gravitational waves}
The confinement/de-confinement phase transition is expected to be first order for $3\le N_F\lesssim 4N$ massless fermions. In this case there can be production of gravitational waves. 
Actually, most likely the transition proceeds in quasi equilibrium \cite{Witten:1984rs,Garcia:2015toa} leading to a very small amplitude. 
Here we entertain the possibility that the transition occurs explosively leading to a larger amplitude. Even so as will show it is quite difficult to obtain 
an observable signal, see \cite{Breitbach:2018ddu} for related work.

If the phase transition completes while the expansion of the Universe is driven by the visible sector we expect a very small power spectrum of gravitational waves, roughly speaking suppressed by a factor $ \xi_0^8$, with respect to the case of an analogous phase transition happening in the visible sector. The reason for this suppression can be understood as follows. If the phase transition in the dark sector happens at the dark nucleation temperature and amount of energy $\rho_{\rm GW}$ is deposited into gravitational waves, the relic abundance today is
\be\label{eq:omega}
\Omega_{\rm gw}=\Omega_{\rm gw}^*  \Omega_\gamma \bigg(\frac{g^{s}_{*}(T_0)}{g^{s}_{*}(T_\Lambda)}\bigg)^{\frac43} \frac{g_*(T_\Lambda)}{g_*(T_0)} \times  \frac{\rho_{\rm tot}(T_n)}{\rho_{R}(T_\Lambda)}\,.
\ee
Where $T_\Lambda$ is the SM temperature after the phase transition, and $g^{s}_*(T)=g^{s}_{*,\rm SM}(T) + \xi^3 g_{*D}(T)$ and $g_*(T)=g_{*,\rm SM}(T) + \xi^4 g_{*D}(T)$. If reheating is instantaneous, by conservation of energy $\rho_{\rm tot}(T_n)=\rho_R(T_\Lambda)$, where $\rho_R$ includes all the relativistic contributions to the energy density $\rho_R=\rho_{\rm SM}+\rho_{D}$.  The expression for $\Omega_{\rm gw}^*$ depends on the production mechanism for gravity waves, by focusing for example on bubble collision contribution we get
\be
\Omega_{\rm gw}^*\big|_{\rm vacuum}\approx \bigg(\frac{H(T_n)}{\beta}\bigg)^{2} \frac{L_h^2}{(\rho_R + L_h)^2}\bigg|_{T_n} \approx \bigg(\frac{H(T_n)}{\beta}\bigg)^{2}\, \frac{900 \frac{L_h^{2}}{T_{D,n}^8}}{\pi^4 g_{SM,*}^2} \frac{\xi_0^8}{(1 + \frac{\xi_0^4}{g_*} (g_D+\frac{30}{\pi^2} \frac{L_h}{T_{D,n}^4} ))^2}\,.  
\ee
Where in the second equality we have estimated the latent heat $L_h= T_{D,c}^4$, with $T_{D,c}$ the critical temperature.
Without substantial supercooling $T_{Dn}\approx T_{D,c}$, therefore we see that $\Omega_{\rm gw}^* \approx \xi_0^8$ and therefore completely negligible.

The only exception to this intrinsic suppression is to explore models with supercooling $T_{Dn}\ll T_{D,c}$ (see for example \cite{Ellis:2019oqb} for supercooling in the visible sector). A phase of supercooling depletes exponentially $\rho_R$ and makes $L_h$ dominant in the above formula, maximizing $\Omega_{\rm gw}^*$. In this scenario after the phase transition, only the dark sector is populated, leading naturally to models with $\xi\to \infty$. As we discussed in section \ref{sec:abundance}, for large $\xi$, the models are viable only if pions decay fast enough to the SM.
In this case, however, the computation of today's abundance has to be revised to take into account the dilution coming from entropy injection due to pion decays.

\begin{figure}[t]
\centering
\includegraphics[width=0.8\linewidth]{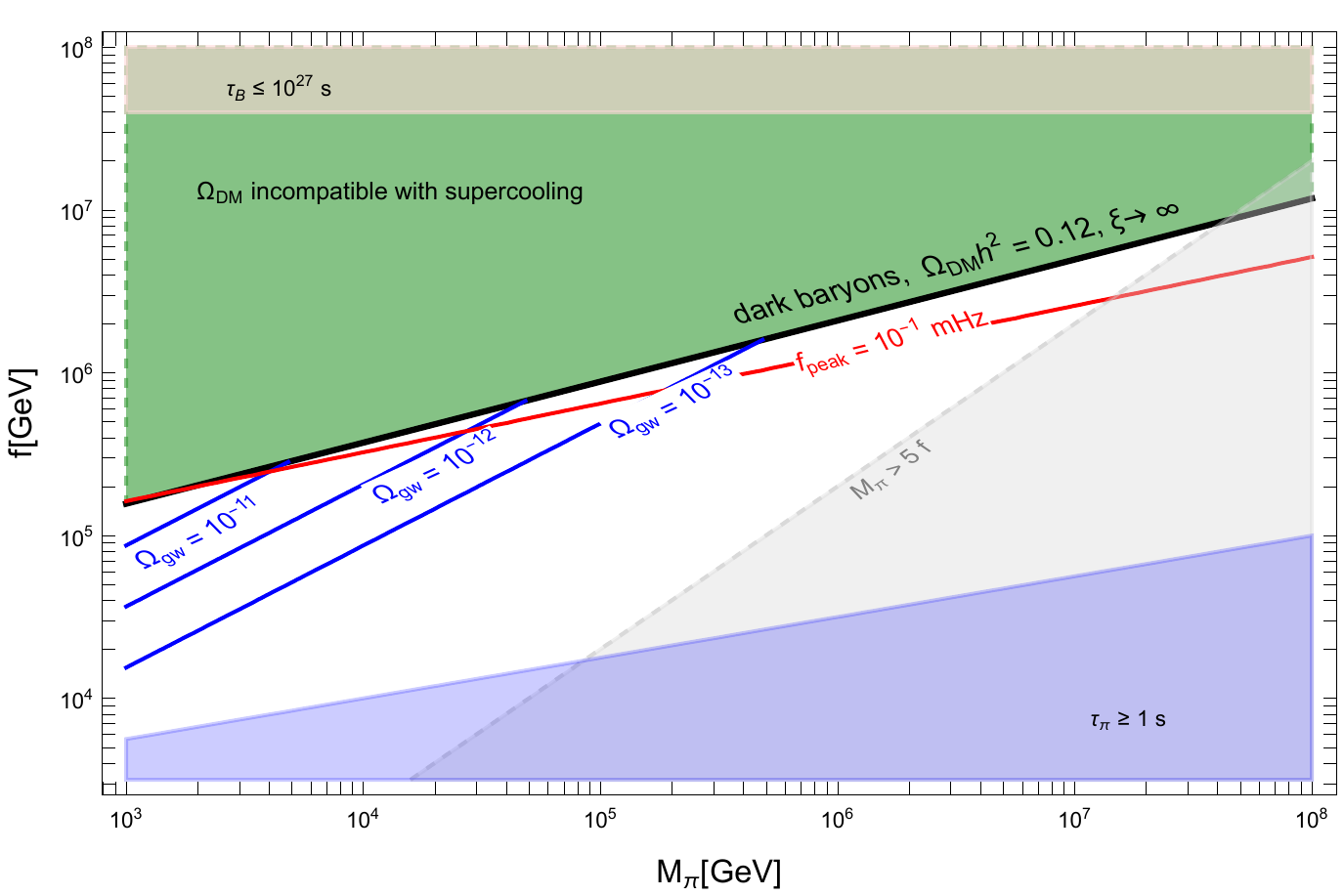}
\caption{\label{fig:gravity}\it Gravity wave amplitude and frequency for the case of unstable pions. Blue and red isolines correspond to the amplitude and frequency of the gravitational waves respectively (assuming supercooling $\beta/H\sim O(1)$). In this region of parameter space the largest signal is achieved when the dark sector is initially at higher temperature than the visible one: the extreme scenario where DM abundance is reproduced starting from an empty visible sector ($\xi\to \infty$) is given by the solid black line.  The points where blue and black lines intersect are predictions for the amplitude of the gravitational wave spectrum with the constraint on DM abundance. }
\end{figure}

\paragraph{Supercooled dark QCD with unstable pions: DM producing the SM}~\\
Let us now consider supercooled phase transitions, with $T_{Dn}\ll T_{D,c}$. When the phase transition completes the dark pions dominate the energy budget of the Universe and then they decay to the SM, reheating it at a temperature $T_{R,\pi}$.
The energy density of gravitational waves at production, $\rho_{\rm gw}$ redshifts as $a^{-4}$, giving today $\rho_{\rm gw,0} =\rho_{\rm gw} (a_*/a_0)^4$. If the reheating of the SM happens instantaneously after the phase transition, then eq.~\eqref{eq:omega} applies. However, in our secluded case the decay to the SM is not fast and we have to take it into account for the computation of the scale factor. Schematically we identify three stages of evolution, $i)$ relativistic pions; $ii)$ matter domination due to pion abundance up to pion decay, $iii)$ standard cosmological evolution starting from a reheating temperature $T_{R,\pi}$ (instantaneous entropy injection). Therefore we approximately decompose the redshift evolution as
\be\label{eq:redshift}
\frac{a_*}{a_{\rm NR}}= \bigg(\frac{\rho_{\rm NR}}{\rho_{\pi}^*}\bigg)^{\frac14},\quad \quad \frac{a_{\rm NR}}{a_{\Gamma}}= \bigg(\frac{\rho_{\rm SM}(T_{R,\pi})}{\rho_{\rm NR}}\bigg)^{\frac13},\quad \quad \frac{a_\Gamma}{a_{0} }= \bigg(\frac{T_0}{T_{R,\pi}}\bigg)\bigg(\frac{g_0^s}{g_R^s}\bigg)^{\frac13}\,,
\ee
where $\rho_\pi^*$ is given by eq.~\eqref{energy-pions} and $T_{R,\pi}$ is the SM (reheating) temperature after the injection of the entropy of the dark sector.  
Assuming instantaneous decay of pions at $H_R\equiv 0.33 \sqrt{g_{*}}T_{R,\pi}^2/\Mpl =\Gamma_\pi$ the overall redshift can be cast in the form,
\begin{equation}
\label{eq:redshift2}
\frac{a_*}{a_0}= \bigg(\frac{a_{\rm NR}}{a_*}\bigg)^{\frac 1 3}\bigg(\frac{3 \Mpl^2 \Gamma_\pi^2}{\rho_\pi^*}\bigg)^{\frac13} \bigg(\frac{T_0}{T_{R,\pi}}\bigg)\bigg(\frac{g_0^s}{g_R^s}\bigg)^{\frac13}\,.
\end{equation}
Here we assume that the Hubble parameter is dominated by the dark sector until pion decay. With supercooling and instantaneous reheating of the dark sector after completion of the PT $\rho_\pi^*\sim \Lambda^4$. Using eq. (\ref{eq:redshift2}) the abundance reads,
\be
\Omega_{\rm gw}^{\pi-\rm decay} = \Omega_\gamma \frac{\rho_{\rm gw}}{\rho_\pi^*} \,\bigg(\frac{g^{s}_{*}(T_0)}{g^{s}_{*}(T_{R,\pi})}\bigg)^{\frac43} \frac{g_*(T_{R,\pi})}{g_*(T_0)}\,\times \bigg(\frac{a_{\rm NR}}{a_*}\bigg)^{\frac43}\bigg(\frac{3\Mpl^2 \Gamma_\pi^2}{\rho_\pi^*}\bigg)^{\frac13}\,
\ee
that holds when $(a_{\rm NR}/a_*)^{\frac43}(3\Mpl^2 \Gamma_\pi^2/\rho_\pi^*)^{\frac13}$ is smaller than one. The peak frequency is also affected by the dilution, becoming smaller
\be
f_{\rm peak}=3.8 \times10^{-6}\, \mathrm{Hz} \, \frac{f_*}{H_*} \bigg(\frac{g_{*}(T_{R,\pi})}{106.75}\bigg)^{\frac16} \bigg(\frac{T_{R,\pi}}{100\GeV}\bigg) 
\times \bigg(\frac{a_{\rm NR}}{a_*}\bigg)^{\frac13} \bigg(\frac{\rho_\pi^*}{3\Mpl^2 \Gamma_\pi^2}\bigg)^{\frac16}\,.
\ee
The numerical values of the above two quantities are shown in figure \ref{fig:gravity}, where we also show the baryon DM abundance compatible with supercooling $(\xi\to \infty)$. 

 As shown in Fig.~\ref{fig:gravity} maximal peak frequency  in the supercooled scenario is $\sim 10^{-1}$ mHz. At such frequencies the maximal amplitude of GWs $\sim 10^{-11}$ could be eventually tested with LISA~\cite{Caprini:2015zlo}. It is worth noting however, at these frequencies and amplitudes stochastic astrophysical foregrounds also exist. Most notably from the mergers of compact objects such as neutron star- neutron star~\cite{Rosado:2011kv} and white dwarf - white dwarf binaries~\cite{Farmer:2003pa,Ruiter:2007xx}. Indeed, to probe our scenario (or any other similar scenario) one has to identify and subtract the astrophysical foregrounds. There are two distinct sources of foregrounds at GW frequencies $\mathcal{O}$(mHz), galactic compact binaries and extra-galactic ones, respectively. It is thought that the galactic component of the foregrounds could be subtracted, however, the extra-galactic binary mergers are thought to contribute to the irreducible background or so called confusion noise~\cite{1998A&A...336..786K,2010PhRvD..82b2002A,2014PhRvD..89b2001A}. Finally, we remark that  foreground subtraction appears to be experimentally challenging even at frequencies $\sim$ Hz which is studied in great detail in refs.~\cite{Regimbau:2016ike,Sachdev:2020bkk}. Considering all the current experimental challenges, supercooled phase transitions could perhaps be tested given a far future experimental break through.

In this section we have made a preliminary assessment of GW signals in the most optimistic case of super-cooled phase transition followed by dilution due to  the decay of massive dark pions. A positive detection of these GWs could point towards new physics realised by models which are classically conformal at high energies. We leave a more dedicated study of GW signals for future work.

\subsection{Phenomenological summary}

From the previous discussion we have identified three possible scenarios that provide the correct relic abundance of dark-baryons and -pions. In this subsection we summarize our findings and comment on the novelty in each of the scenarios. 
The overall parameter space is shown in figure \ref{fig:abundance}.

\paragraph{Baryon DM + Pion DM}~\\
If the dark pions are lighter than GeV, they can be cosmologically stable. In this branch, both baryons and pions are DM, and the mass scale varies significantly with the value of $\xi$. The baryons are the dominant component of DM as long as the pions are lighter than $M_\pi < 100\, \mathrm{MeV} (f/10^5\GeV)^6$. This scenario is represented in the left region of figure \ref{fig:abundance}. For $\xi< 10^{-3}$ the DM abundance is dominated by the baryons (horizontal solid lines in fig. \ref{fig:abundance}), and it corresponds to the darkest scenario, with basically no observable effects.
For moderate value of $\xi$ instead and for DM dominated by the pions, there could be visible effects in structure formation, both from the Bullet Cluster and Ly-$\alpha$ constraints. The latter strongly disfavor secluded dark sectors with $\xi\approx 1$. It is interesting for example to consider the case of an initial value of $\xi_0=1$, such a scenario with equal initial temperatures is realized if renormalizable interactions exist between the dark sector and the SM, most simply if there exist heavy fermions charged under both the SM and the dark sector.
Note that from eq. (\ref{eq:xiprime}) after the phase transition the dark sector temperature slightly increases.
For example for $N=N_F=3$ one finds $\xi=1.85$ (while $N=3$ with $N_F=2$ gives $\xi=2.35$). Such values of $\xi$ are grossly excluded by structure formation.

\paragraph{Baryon DM with ultra-light pions}~\\
When the dark pions are so light that they cannot be DM, they behave as radiation at the BBN and CMB epoch, therefore they are subject to the bound from $\Delta N_{\rm eff}$. This region, where DM is made entirely by the baryons is the leftmost part of figure \ref{fig:abundance}. For moderate $\xi$ this region is constrained by the value of the number of relativistic degrees of freedom. In the plot we show the expected bound from CMB stage IV, in dark green.

\paragraph{Baryon DM with fast-decaying pions}~\\
For larger dark pion mass, pions are unstable, although they could be sufficiently long lived to modify the baryon relic abundance through late entropy injections (still with $\tau_\pi < 1 s$ to avoid BBN constraints). This region is on the top-right part of figure \ref{fig:abundance}. Since DM baryon are heavy, there is no constraint from self-interactions in this branch of the parameter space. The relic abundances isolines of figure \ref{fig:abundance} are of two types in this region: horizontal lines with no dependence on $M_\pi$ and oblique lines with dependence on $M_\pi$. The former correspond to small values of $\xi$, that are insufficient to achieve a large entropy injections, while the latter to moderate and large values of $\xi$. When $\xi$ is sizable, the DM abundance of baryon does not depend anymore on $\xi$: this is the limiting case where the SM is extremely cold initially and it originates entirely from the dark sector (cfr. eq.~\eqref{eq:baryonDMmass}). As discussed in the previous subsection it is the region of parameter space where we can expect a signal in gravitational waves, albeit a very tiny one and possibly unobservable. The most optimistic predictions for the amplitude and frequency of the gravitational waves are shown in figure~\ref{fig:gravity}, which is a zoomed-in version of the upper right region of figure~\ref{fig:abundance}.

\section{Conclusions}\label{sec:conclusions}

If DM is part of a truly dark sector, with no sizable interactions with the SM, the experimental chances to have a glimpse of the nature of DM are dim. However, while being clearly a nightmare scenario, this possibility cannot be merely discarded. A strong theoretical motivation to study secluded dark sectors is that they elegantly provide cosmologically stable DM candidates without ad hoc assumptions. The seclusion is automatically realized when the dark sector is a non-abelian gauge theory with fermions that are singlet under the SM. This in turn  gives rise to interactions with potentially interesting effects for cosmology.

In this work we studied a QCD-like dark sector, connected to the SM only through gravitationally suppressed interactions. This leads to dark baryon and dark pion DM candidates in different regions of parameter space. Depending on the production mechanism the dark sector has a different temperature from the visible sector and this determines the DM relic abundance and phenomenology. This simple example already generates reach and non-trivial dynamics: an early phase of dark radiation, then a confinement (chiral symmetry breaking) phase transition to a dark sector with two mass scales, the baryon and pion mass.

In the context of gauge theories with fermions a dimension-5 operator through the Higgs portal has a dramatic impact even when suppressed by the Planck scale.
This boosts the production of the dark sector through freeze-in  and it allows the pions to decay. The latter effect can modify the cosmological history of the Universe with an early phase of matter domination, and severely constrains the scenario if the pions decay at late times. Contrary to the pure glue scenario \cite{Redi:2020ffc}, the existence of pions that are lighter than the confinement scale opens new phenomenological avenues.

Quite remarkably we have shown that gravitationally coupled dark QCD is quite constrained through a combination of constraints from CMB, BBN, structure formation and self-interactions
and can be further tested with future observations. The constraints depend on the initial temperature of the dark sector. If the dark sector was originally in thermal contact with the SM 
only a small region of parameters is allowed where DM is a baryon and pions decay rapidly. 

Two regions of parameters are currently allowed.
If the pions are lighter than GeV they can make up all the DM and be as light as the mass scale currently tested with Ly-$\alpha$ forest observations.  
We have carried out a very preliminary study of the impact of light pion DM on structure formation, emphasizing both the role of free-streaming and self-interactions. It turned out that both effects gives parametrically the same model dependence on the matter power spectrum. When the temperature of the dark sector is equal to the SM pion DM would have mass around $\mathrm{KeV}$ and this is grossly excluded by structure formation.

On the contrary, when the dark pions are heavier than the Higgs mass they decay  before BBN but they can be sufficiently long lived to realize an early phase of matter domination. Upon decay to the SM such a phase ends with a large entropy injection into the SM plasma diluting DM abundance. This leads to baryon DM with mass 100 TeV or larger.
In such a scenario the energy budget of the Universe at the dark QCD phase transition might be dominated by the dark sector, opening up the possibility to have signals of gravitational waves from the first order confinement phase transition, albeit their amplitude and peak frequency are diluted by the entropy injection. Our preliminary study indicates that even in the most optimistic case of sizable supercooling (that is unlikely in QCD-like theories), the peak amplitude and frequency are about $\Omega_{GW} \sim 10^{-11}$ and $\mathcal{O}$(mHz), respectively. Allowing for a faster decay of the pions might lead to larger observable gravity wave signals. 

The exploration of truly dark sectors can be pursued in several future directions. We plan to explore more general portal interactions between the visible and the dark sector, employing the formalism of CFTs, and generalize the discussion of freeze-in in section \ref{sec:production}. Dark sectors with a tiny connection to the SM can provide an early phase of matter domination terminated by the decay to the SM via irrelevant operators. Since this can have an impact for gravitational waves, we reserve to explore this possibility in greater detail in future work. For baryon DM indirect detection signal of decaying pions deserve further study. Finally for pion DM a more detailed study of effects on structure formation due to free-streaming/self-interactions is required in some regions of parameters.

{\small
\subsubsection*{Acknowledgments}
This work is supported by MIUR grants PRIN 2017FMJFMW and 2017L5W2PT and the INFN grant STRONG. 
We acknowledge the Galileo Galilei Institute for hospitality during this work. We thank Yann Gouttenoire for pointing out a typo 
in the gravity-wave frequency.}

\appendix

\section{Production of CFT from contact operators}\label{app:hgportal}

In this Appendix we extend the computation in Ref.~\cite{Redi:2020ffc} to the production of a dark sector through the Higgs portal coupling,
\begin{equation}
\frac 1 {\Lambda_{\rm UV}^{d-2}}|H|^2 {\cal O}\,,~~~~~~~~~~~~~[{\cal O}]=d \,.
\end{equation}
We will phrase our formulae for a general operator $\cal{O}$ of the CFT, see \cite{Contino:2020tix} for a recent discussion.  
As a special case they can be applied to perturbative gauge theories with fermions  and gauge fields or conformally coupled scalars. The two point function of ${\cal O}$ in real and Fourier space is given by
\be
\langle \op(x)\op(0)\rangle =\frac{a_\op}{8\pi^4}\frac{1}{(x^2)^d}\,, \quad \langle \op(p)\op(-p)\rangle'= -i\frac{ a_\op}{2\pi^2}\frac{\Gamma(2-d)}{4^{d-1} \Gamma(d)} (-p^2)^{d-2}\,,
\ee
The total cross-section for production of CFT states from two Higgs can be simply obtained through the optical theorem.
In the massless limit $\sigma_{\rm tot}= {\rm Im}{\cal M_{\rm forward}}/s$, where
\begin{equation}
i{\cal M}_{\rm forward}= -\frac{4}{\Lambda_{\rm UV}^2} \langle O(p) O(-p)\rangle'\,.
\end{equation}
By taking the imaginary part one finds
\begin{equation}
\sigma_{\mathrm{HH \to CFT}}= \frac{4 a_\op}{\pi^{3/2}} \frac {\Gamma(d+1/2)}{\Gamma(d-1)\Gamma(2d)} \frac{s^{d-3}}{\Lambda_{\rm UV}^{2d-4}}\,.
\label{eq:sigmaH2CFT}
\end{equation}
With this we can compute the collisional term in the Boltzmann equation, which can be cast into the following form \cite{Redi:2020ffc}
\begin{equation}
\frac {C(t,p)}{E}\approx \frac  {e^{-p/T}}{512 \pi^3 p^2} \int ds \int_{\frac {s} {4 p}}^{\infty} dp_3 \frac 2 s e^{-p_3/T}  16\pi \sigma(s)=
T \frac  {e^{-p/T}}{16 \pi^2 p^2} \int ds s e^{-s/(4 p T)} \sigma(s)\,.
\end{equation}
Inserting (\ref{eq:sigmaH2CFT}), we obtain the expression
\begin{equation}
\frac {C_{\rm HH \to CFT}}{E}= a_\op \frac {p^{d-3}e^{-\frac p T}}{2\pi^3\Gamma(d)} \frac{T^d}{\Lambda_{\rm UV}^{2d-4}}\,.
\label{eq:collision}
\end{equation}
This can be used to determine both the thermally averaged cross-section and to find a solution for the phase space distribution by direct integration of the Boltzmann equation.
We can compute the thermally averaged cross-section as
\begin{equation}
2\gamma\equiv n_{\rm eq}^2 \langle \sigma_{\mathrm{HH\to CFT}} v\rangle= \int \frac{d^3 p}{(2\pi)^3} \frac {C_{\rm SM+SM \to CFT}}{E}=\frac{a_{\cal O}}{4\pi^5} \frac {T^{2d}}{\Lambda_{\rm UV}^{2d-4}} \,,
\end{equation}
where $n_{\rm eq}= g_i T^3/\pi^2$ in agreement with \cite{Gondolo:1990dk}.  This relation then allow us to compute the thermally averaged cross section as in eq.~\eqref{eq:thermalxsec} in section \ref{sec:production}. From eq. (\ref{eq:collision}) we then derive the phase space distribution as a function of momenta and temperature, assuming that production takes place during radiation domination,
\begin{equation}
f(T,p)=\int_{T}^{T_R} \frac {dT'}{T' H(T')} \frac {C(T',\frac{p\,T'} T)}{p}=\frac {a_\op}{\pi^4}\frac{ 3\sqrt{5/2}} {(2d-5)\Gamma(d)\sqrt{g_*}}\bigg(1- \frac {T^{2d-5}}{T_R^{2d-5}}\bigg)\frac {\Mpl}{\Lambda_{\rm UV} }\frac{T_R^{2d-5}}{\Lambda_{\rm UV}^{2d-5}} \frac{p^{d-3}}{T^{d-3}} e^{-p/T}\,.
\end{equation}
From this we can finally compute the number and energy densities
\begin{equation}
n(T)=a_{\cal O} \frac {3\sqrt{5/2}} {2\pi^6(2d-5)\sqrt{g_*}} T^3\left(1-\frac {T^{2d-5}}{T_R^{2d-5}}\right)\frac {\Mpl}{\Lambda_{\rm UV} }\frac{T_R^{2d-5}}{\Lambda_{\rm UV}^{2d-5}} \,,\quad\quad \rho(T)= d \,T n(T)\,.
\end{equation}

\paragraph{Corrections from quantum statistics}
In the previous discussion we neglected quantum statistic for the SM bath. While this is completely negligible in the non-relativistic regime 
it can give a correction in the massless limit. To take this into account  the space-time density of interaction  in the massless limit can be written as \cite{Gondolo:1990dk},
\begin{equation}
2\gamma = \int  \frac{d^3 p_1}{(2\pi)^3}\frac {d^3 p_2}{(2\pi)^3} f_{\rm eq}(E_1) f_{\rm eq}(E_2) \frac{s \sigma}{2 E_1 E_2}\,.
\end{equation}
Where $f_{\rm eq}(x)=(\exp[x]\mp 1)^{-1}$ for Bose/Fermi statistics.
Using,
\begin{eqnarray}
&d^3 p_1 d^3 p_2= 2\pi^2  E_1 E_2 dE_+ dE_- ds\,,~~~~~~~~~~~~~E_{\pm}=E_1\pm E_2\,,\nonumber \\
& |E_-| \le \sqrt{E_+^2-s}\,,~~~~~E_+\ge\sqrt{s}\,,~~~~~~ s\ge 0\,.
\end{eqnarray}
we find,
\begin{equation}
2 \gamma=I_d\frac {a_D} {16 \pi^4} \frac {2^{5-2 \Delta}}{\Gamma[d]\Gamma[d-1]} \frac{T^{2d}}{\Lambda_{\rm UV}^{2d-4}}  \,,~~~~I_d= \int_0^\infty dy\,y^{d-2}  \int_{\sqrt{y}}^{\infty} d x_+  \int_0^{x_+} d x_-  f_{\rm eq}(x_1) f_{\rm eq}(x_2) 
\end{equation}
Numerically we find,
\begin{center}
\begin{tabular}{c||c|c|c|c}
 &  $I_3$ & $I_4$ & $I_5$ & $I_6$\\ 
\hline 
\text{Bose} &  92 & 1799 & 79273 & 6100000\\
\text{Fermi} &  52 & 1377 &  69674 & 5700000 \\
\text{Maxwell} & 64 & 1536 & 73728 & 5800000 \\
\end{tabular}
\end{center}
For gravitational production the relative value is the same as $d=4$ corresponding to an O(10\%) difference.

\subsection{Inflaton scattering}\label{app:inf_scat}
Let us consider gravitational production of the dark sector from inflaton collisions.
As usual the inclusive cross-section is proportional to the imaginary part of the forward amplitude of $\phi\phi \to \phi\phi$.
Using the tree level graviton propagator
\begin{equation}
\langle h_{\mu\nu}(p) h_{\rho \sigma }(-p)\rangle = P_{\mu\nu\rho\sigma} \frac {i}{p^2+i \epsilon}\,,~~~~~~~~~~~~P_{\mu\nu\rho\sigma}= \frac 1 2(\eta_{\mu\rho}\eta_{\nu \sigma}+\eta_{\mu\sigma}\eta_{\nu \rho}-\eta_{\mu\nu}\eta_{\rho \sigma})\,,
\end{equation}
one finds,
\begin{equation}
i {\cal M} = \frac {1}{(p^2)^2} T^{\phi}_{\mu\nu} P^{\mu\nu\rho\sigma} \langle T_{\rho\sigma}(p) T_{\gamma \tau }(-p)\rangle P^{\gamma\tau\alpha\beta} T^{\phi}_{\alpha\beta}\,.
\end{equation}
The two point function of the energy momentum tensor of a CFT is fixed up to an overall normalization.
One finds \cite{Gubser:1997se},
\begin{equation}
\langle T_{\mu\nu}(p) T_{\rho \sigma }(-p)\rangle= \frac{c}{7680 \pi^2}  \left(2\pi_{\mu\nu} \pi_{\rho\sigma}-3\pi_{\mu\rho} \pi_{\nu\sigma}-3\pi_{\mu\sigma} \pi_{\nu\rho}\right)  \log(-p^2)\,,
\end{equation}
where $\pi_{\mu\nu}=\eta_{\mu\nu}p^2-p_\mu p_\nu $ and $c$ is the central charge of the CFT. Since the inflaton is non-relativistic  $p_\mu=(2m,0,0,0)$ the cross-section vanishes identically since $\pi_{\mu\nu}=0$.
This agrees with perturbative computations where the cross-section is controlled by the explicit breaking of Weyl invariance.

\pagestyle{plain}
\bibliographystyle{jhep}
\small
\bibliography{biblio}

\end{document}